\documentclass[twocolumn, tighten, times]{aastex631}
\usepackage{CJK}
\usepackage{graphicx}
\usepackage{epstopdf}
\usepackage{float} 
\usepackage{bm}
\usepackage{subfigure} 
\usepackage{diagbox}

\usepackage{color}
\usepackage{soul}

\DeclareUnicodeCharacter{02BC}{}

\shorttitle{Classification of Cosmos-Dash}
\shortauthors{Yao Dai}

\begin{document}
\begin{CJK*}{UTF8}{gbsn}
\title{The Classification of Galaxy Morphology in H-band of COSMOS-DASH Field: a combination-based machine learning clustering model}
\correspondingauthor{Guanwen Fang}
\email{wen@mail.ustc.edu.cn, xkong@ustc.edu.cn}
\author[0000-0002-4638-0235]{Yao Dai (代瑶)}
\affil{Institute of Astronomy and Astrophysics, Anqing Normal University, Anqing 246133, People's Republic of China; 
\url{wen@mail.ustc.edu.cn}} 

\author[0000-0003-1697-6801]{Jun Xu (徐骏)}
\altaffiliation{Jun Xu and Yao Dai contributed equally to this work}
\affil{Institute of Astronomy and Astrophysics, Anqing Normal University, Anqing 246133, People's Republic of China; 
\url{wen@mail.ustc.edu.cn}} 

\author[0000-0002-0846-7591]{Jie Song (宋杰)}
\affil{Deep Space Exploration Laboratory / Department of Astronomy, University of Science and Technology of China, Hefei 230026, China; \url{xkong@ustc.edu.cn}} 
\affil{School of Astronomy and Space Science, University of Science and Technology of China, Hefei 230026, People's Republic of China}

\author[0000-0001-9694-2171]{Guanwen Fang (方官文)}
\affil{Institute of Astronomy and Astrophysics, Anqing Normal University, Anqing 246133, People's Republic of China; 
\url{wen@mail.ustc.edu.cn}} 

\author[0000-0002-5133-2668]{Chichun Zhou (周池春)}
\affil{ School of Engineering, Dali University, Dali 671003, People's Republic of China} 

\author{Shuo Ba (巴朔)}
\affil{ School of Engineering, Dali University, Dali 671003, People's Republic of China} 

\author[0000-0003-3196-7938]{Yizhou Gu (顾一舟)}
\affil{School of Physics and Astronomy, Shanghai Jiao Tong University, 800 Dongchuan Road, Minhang, Shanghai 200240, People's Republic of China}

\author[0000-0001-8078-3428]{Zesen Lin (林泽森)}
\affiliation{Department of Physics, The Chinese University of Hong Kong, Shatin, N.T., Hong Kong S.A.R., China}

\author[0000-0002-7660-2273]{Xu Kong (孔旭)}
\affil{Deep Space Exploration Laboratory / Department of Astronomy, University of Science and Technology of China, Hefei 230026, China; \url{xkong@ustc.edu.cn}} 
\affil{School of Astronomy and Space Science, University of Science and Technology of China, Hefei 230026, People's Republic of China}

\begin{abstract}
By applying our previously developed two-step scheme for galaxy morphology classification, we present a catalog of galaxy morphology for H-band selected massive galaxies in the COSMOS-DASH field, which includes 17292 galaxies with stellar mass $M_{\star}>10^{10}~M_{\odot}$ at $0.5<z<2.5$. The classification scheme is designed to provide a complete morphology classification for galaxies via a combination of two machine-learning steps. We first use an unsupervised machine learning method (i.e., bagging-based multi-clustering) to cluster galaxies into five categories: spherical (SPH), early-type disk (ETD), late-type disk (LTD), irregular (IRR), and unclassified (UNC). About 48\% of galaxies (8258/17292) are successfully clustered during this step. For the remaining sample, we adopt a supervised machine learning method (i.e., GoogLeNet) to classify them, during which galaxies that are well-classified in the previous step are taken as our training set. Consequently, we obtain a morphology classification result for the full sample. The t-SNE test shows that galaxies in our sample can be well aggregated. We also measure the parametric and nonparametric morphologies of these galaxies. We find that the S\'{e}rsic index increases from IRR to SPH and the effective radius decreases from IRR to SPH, consistent with the corresponding definitions. Galaxies from different categories are separately distributed in the $G$--$M_{20}$ space. Such consistencies with other characteristic descriptions of galaxy morphology demonstrate the reliability of our classification result, ensuring that it can be used as a basic catalog for further galaxy studies.
\end{abstract}

\section{Introduction} \label{sec:intro}
Galaxy morphology and how it evolves with time are crucial in understanding the assembling history and evolution of galaxies. Various galaxies exhibit different features (e.g., budge, spiral arm, bar, and tidal tail). By visual inspection of about 400 galaxy photographic images, \cite{Hubble_1926} presented a systematic study of galaxy morphology, which found that galaxies can be mainly divided into four categories (i.e., Spiral, Lenticular, Elliptical and Irregular), and proposed the Hubble sequence scheme. These galaxy morphology categories are then found to be connected to other physical parameters. For instance, color, gas content, star formation rate, stellar mass and environment \citep{Schawinski2014,Gu2018,Kauffmann2003,Kauffmann2004,Kawinwan2017,Dressler1980,Lianou2019,Omand2014}. The diverse properties of galaxies in different morphology categories may imply different evolution paths. To understand galaxy evolution, the key is to obtain reliable classification results of galaxies at each epoch in the universe.

There are several ways to derive the morphological type of galaxies. Visual inspection is a commonly used direct way since \cite{Hubble_1926} and is still widely used in some projects. The Galaxy Zoo is a significant project of visual inspection that gets nearly half a million volunteers involved. In the project, the morphological type of each source is voted on by a certain number of volunteers by recognizing features in the image \citep{walmsleyIdentificationLowSurface2019,simmonsGalaxyZooQuantitative2017a}. This method shows good robustness when signal-to-noise ratio and resolution change between images, but in the meanwhile prohibitively time-consuming. Apart from the visual inspection, the multidimensional morphological parameter space is a practical tool in galaxy morphology classification when taking an empirical cutoff. For example, some non-parametric statistics (e.g., concentration, asymmetry, clumpiness, $M_{20}$, and the Gini coefficient) are designed to describe the characteristics of galaxies \citep{abrahamNewApproachGalaxy2003,conseliceDIRECTMEASUREMENTMAJOR2003,lotzNewNonparametricApproach2004}. Galaxy morphology could be distinguished within the parameter space \citep{conseliceDIRECTMEASUREMENTMAJOR2003,lotzNewNonparametricApproach2004}. These parameters describe the certain morphological features of galaxies quantitatively, but drop much information in the image and thus may lead to failure in classification.

In recent years, machine-learning technology such as the convolutional neural network (CNN) has been applied to derive galaxy morphology
automatically \citep{dielemanRotationinvariantConvolutionalNeural2015,huertas-companyCATALOGVISUALLIKEMORPHOLOGIES2015a,walmsleyIdentificationLowSurface2019}. By taking advantage of the abundant information in the raw image, the CNN method has been applied to SDSS \citep{dielemanRotationinvariantConvolutionalNeural2015} and CANDELS images \citep{huertas-companyCATALOGVISUALLIKEMORPHOLOGIES2015a}. Since CNN is a supervised machine learning (SML) method, it highly depends on the prior information from the training set to simulate human perceptions. Meanwhile, unsupervised machine learning (UML) is another kind of machine-learning technology, which does not need a pre-labeled training set. It clusters galaxies by the characteristics of the image itself, even if the machine does not understand the galaxy features. As a result, it is widely used in morphology analysis in the era of big data survey \citep{ralphRadioGalaxyZoo2019,galvinCataloguingRadioskyUnsupervised2020}. Generally, UML methods work in two steps: (1) extract features from the raw image, and (2) cluster galaxies by similar features.

Various UML methods have been designed in practice. For example, \cite{hockingAutomaticTaxonomyGalaxy2018} and \cite{chengIdentifyingStrongLenses2020} extracted features using the growing neural gas algorithm \citep{article} and cluster the galaxies with the hierarchical clustering technique. 
The convolutional autoencoder (CAE; \citealt{inproceedings}) is another effective technique for extracting image features. 
\cite{2022AJ....163...86Z} applied CAE and a Bagging-based multi-clustering model to cluster CANDELS images and obtained a reliable classification result with a cost of rejecting a certain fraction of disputed sources that reach no agreement in the voting of the bagging method. Later, by adopting the classification result of \cite{2022AJ....163...86Z} as a training set, \cite{2023AJ....165...35F} used an SML method to classify the rejected sources in \cite{2022AJ....163...86Z}. Thus, by combining the UML and SML methods, we are able to classify the galaxy sample into different morphological categories entirely.

COSMOS-DASH is the largest near-infrared (NIR) survey using HST/WFC3, which could help us study the morphology of galaxies at redshift $0.5<z<2.5$, where the rest-frame optical emission shifts into NIR. In this paper, we apply both UML (i.e., CAE \& bagging-based multi-clustering algorithm) and SML (i.e., GoogLeNet) methods to massive galaxies in the COSMOS-DASH field to get reliable and complete morphology classification result.

The paper is organized as follows. Section \ref{sec:data} describes the COSMOS-DASH survey and the sample we used. We introduce the UML method and the GoogLeNet model in Section \ref{sec:method}. In Section \ref{sec:inspection}, We present the test of the classification results in the galaxy parameter space and provide a catalog. Finally, a conclusion will be given in Section \ref{sec:results}.

\section{DATA AND SAMPLE SELECTION} \label{sec:data}
\subsection{COSMOS-DASH} \label{subsec:COSMOS-DASH}

Wide-field NIR survey is vital in studying galaxies at high redshift, where the rest-frame optical emissions shift into the NIR bands. Various projects are conducted by ground-based facilities (e.g., NMBS, \citealt{whitakerNEWFIRMMEDIUMBANDSURVEY2011}; UltraVISTA, \citealt{mccrackenUltraVISTANewUltradeep2012a,2013ApJS..206....8M}) and space facilities (e.g., HST, JWST). For the HST, it is hard to balance resolution, depth, and area for observations. To obtain high-resolution deep-field images, observations should be limited to a tiny field of view, which makes large-scale deep-field NIR sky surveys very difficult. Drift and Shift \citep[DASH;][]{2017PASP..129a5004M} is an efficient technique for wide-field observation with HST. With the DASH technique, \cite{mowlaCOSMOSDASHEvolutionGalaxy2019} present a wide-field NIR survey of the COSMOS field, which is also named COSMOS-DASH. It is taken with 57 DASH orbits in the F160W filter of WFC3 and covers an area of 0.49 $\rm deg^2$ (0.7 $\rm deg^2$ when combined with archival data), which is much larger than the CANDELS field. Since the exposures are around 300s per pointing,  the $5\sigma$ source depth of the image is $H_{160}=25.1$ ABmag. The COSMOS-DASH field is centered at R.A.=10:00:28.6, decl.=+02:12:21.0 and contains $50000\times 50000$ pixel (with $0\farcs1$ per pixel). The final mosaic of the image is available from the COSMOS-DASH website.\footnote{\url{https://archive.stsci.edu/hlsp/cosmos-dash/}} 

\subsection{UVISTA Catalog} \label{subsec:catalog}
To obtain stellar mass and other physical parameters of galaxies, we select our sample from the UltraVista $K_s$-selected catalog \citep{2013ApJS..206....8M}, which is based on an early release of the NIR data (UltraVista DR1). The catalog was generated by the PSF matching images in 30 filters. They divided the UltraVISTA into nine separate pointings depending on the layout of the COSMOS Suprimecam. Moreover, PSF matching was done separately in each of the nine fields to optimize any field-to-field PSF variations. They choose the UltraVista $K_s$ band as the selection ban and reach a depth of $K_{s, tot} = 23.4$ ABmag at 90\% completeness. The photometric redshift of each galaxy in the catalog was determined by fitting the spectral energy distributions (SEDs) within $0.1-24\ \mu$m by using the EAZY code \citep{Brammer_2008}. The photometric redshift had been tested by the spectral redshift of galaxies from COSMOS.
Moreover, the rest-frame colors were extracted from the outputs of the EAZY code. Along with the redshifts, they also fit the galaxy's SEDs to the \cite{bruzualStellarPopulationSynthesis2003} stellar population synthesis models to derive stellar mass with the FAST code \citep{kriekULTRADEEPNEARINFRAREDSPECTRUM2009}. During the fitting process, they assumed a \cite{chabrierGalacticStellarSubstellar2003} initial mass function, an exponentially declining star formation history, a \cite{calzettiDustContentOpacity2000} dust attenuation curve, and solar metallicity. Although the depth of this catalog is only $K_{s, tot} = 23.4$ mag, it is deep enough to select the massive galaxies analyzed in this paper.

\subsection{Selection of Galaxies for Analysis} \label{subsec:sample selection}

This paper aims to derive the morphology classification result of massive galaxies in the COSMOS-DASH field. We adopt the UltraVISTA/$K_s$ selected catalogs and HST/F160W images from the COSMOS-DASH survey. We study the massive galaxies with $M_{\star} > 10^{10}M_{\odot}$ at $0.5<z<2.5$, which are bright enough to derive reliable morphologies. Since there are a few bright stars in the field, we also set the criterion $use = 1$ to ensure reliable stellar mass estimation. The flag means the objects 1) are not too faint (i.e., $K_s<23.5$); 2) are a galaxy rather than a star; 3) are not near a bright star; 4) are only missing a few filters of data, so their photometric redshift and stellar population fits are reliable. Finally, 17292 galaxies are selected in our final sample after removing images with bad pixels.

\section{The Method for Morphological Classification} \label{sec:method}

In this section, we present the scheme we use to classify the morphology of these galaxies (as shown in Figure~\ref{fig:fig01}). In our previous work, \cite{2022AJ....163...86Z} developed a UML method to classify galaxies with similar morphologies in deep field surveys automatically. The UML method consists of two steps:
\begin{figure}[htbp!]
\centering
\includegraphics[scale=0.23]{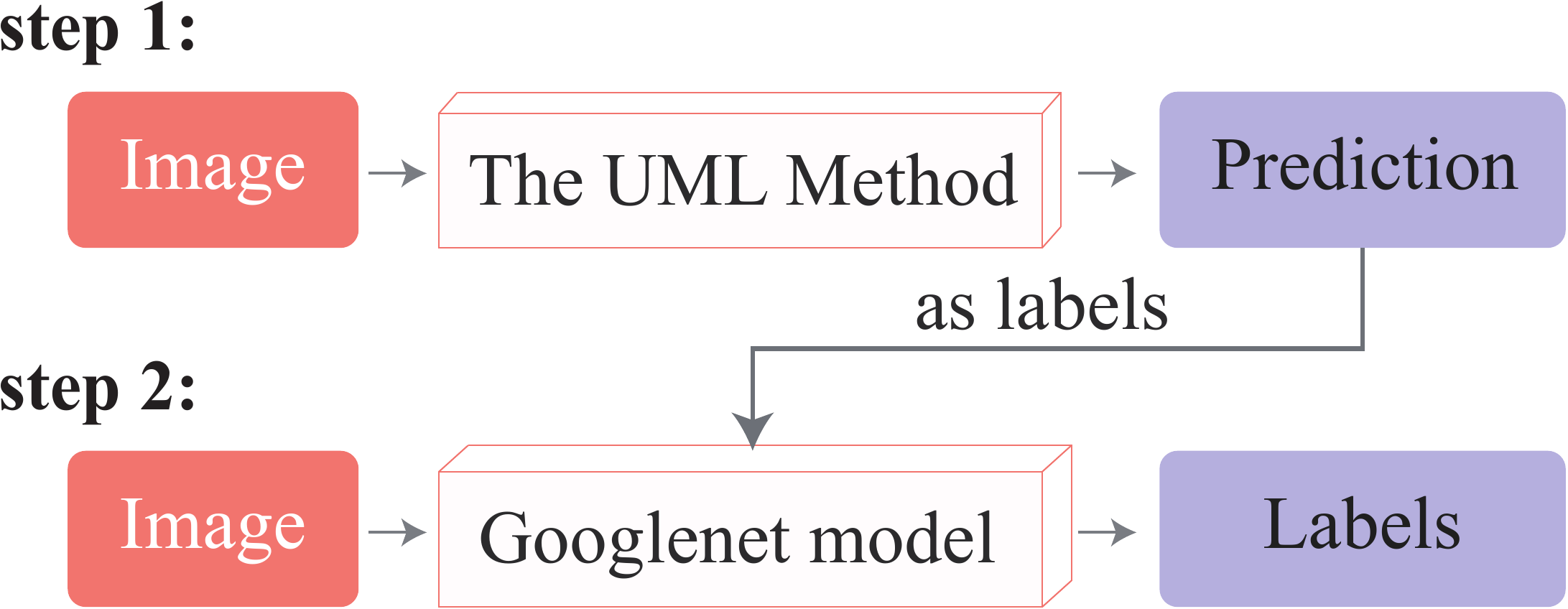}  % 图片路径
\caption{Framework of the combination-based machine learning clustering model. Step 1: Unsupervised clustering of images was carried out to obtain labels; 
 Step 2: Supervised classification is carried out on the unclassified images to obtain the complete classification result of the data set so that the sample data can be fully utilized.}  % 图片标题
\label{fig:fig01}    % 标签，用来引用
\end{figure}

(1) Use CAE to compress the dimensions of the original data and extract the features; (2) Based on the bagging clustering method, guaranteed galaxies with similar characteristics are classified into one group. After discarding the galaxies with inconsistent voting results, the remaining galaxies were nicely grouped into 100 groups. Then by visual classification, 100 types of galaxies with similar features are classified into five categories, including spherical (SPH), early type disk (ETD), late-type disk (LTD), irregular disk (IRR), and unclassified (UNC). As demonstrated in \cite{2022AJ....163...86Z}, among the three models used by the bagging-based multi-clustering method, GoogLeNet has a high classification efficiency in the morphology classification of galaxies in a deep field. Therefore, following \cite{2023AJ....165...35F}, we use the GoogLeNet model as our SML algorithm to classify the remaining sources that are discarded by the UML method so that we can fully utilize the sample data and realize the purpose of complete classification. The SML method consists of two steps: (1) The GoogLeNet model is trained by adopting galaxies successfully classified by the UML method as the training set. (2) The trained GoogLeNet model is applied to classify the discarded sources in the UML step.

\subsection{Data Preprocessing}
\label{subsec:data_prepro}

\begin{figure*}[htbp]
\centering
\includegraphics[scale=0.55]{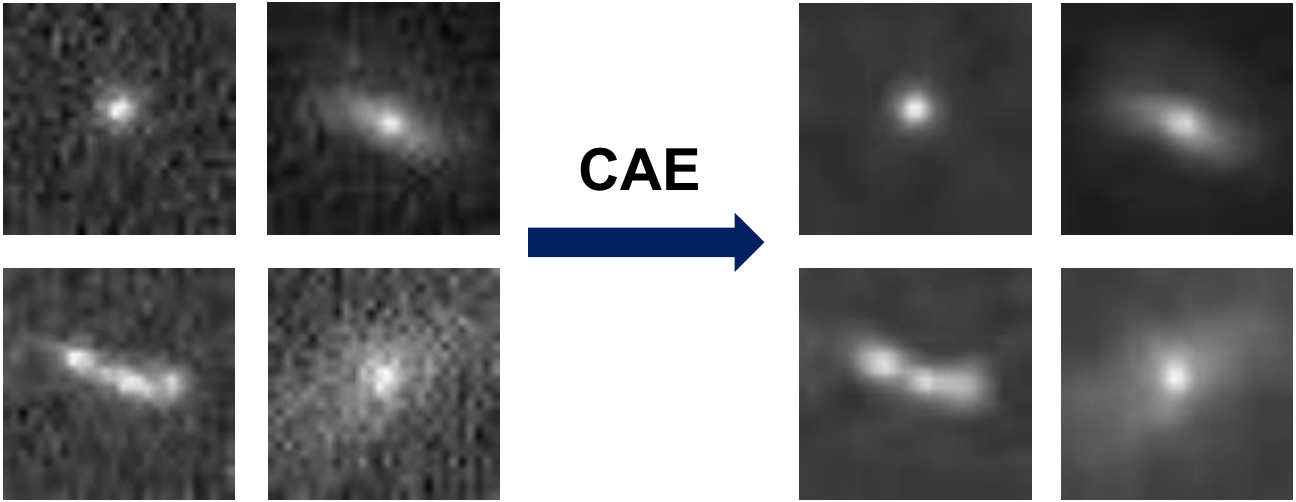}  % 图片路径
\caption{Comparisons between images of the original inputs (left) and the reconstructed ones after the CAE processing (right). The usage of CAE preserves most of the original morphological features of galaxies and eliminates unnecessary background noise meanwhile.}  % 图片标题
\label{fig:fig02}    % 标签，用来引用
\end{figure*}
\begin{figure*}[htbp]
\centering
\includegraphics[scale=0.4]{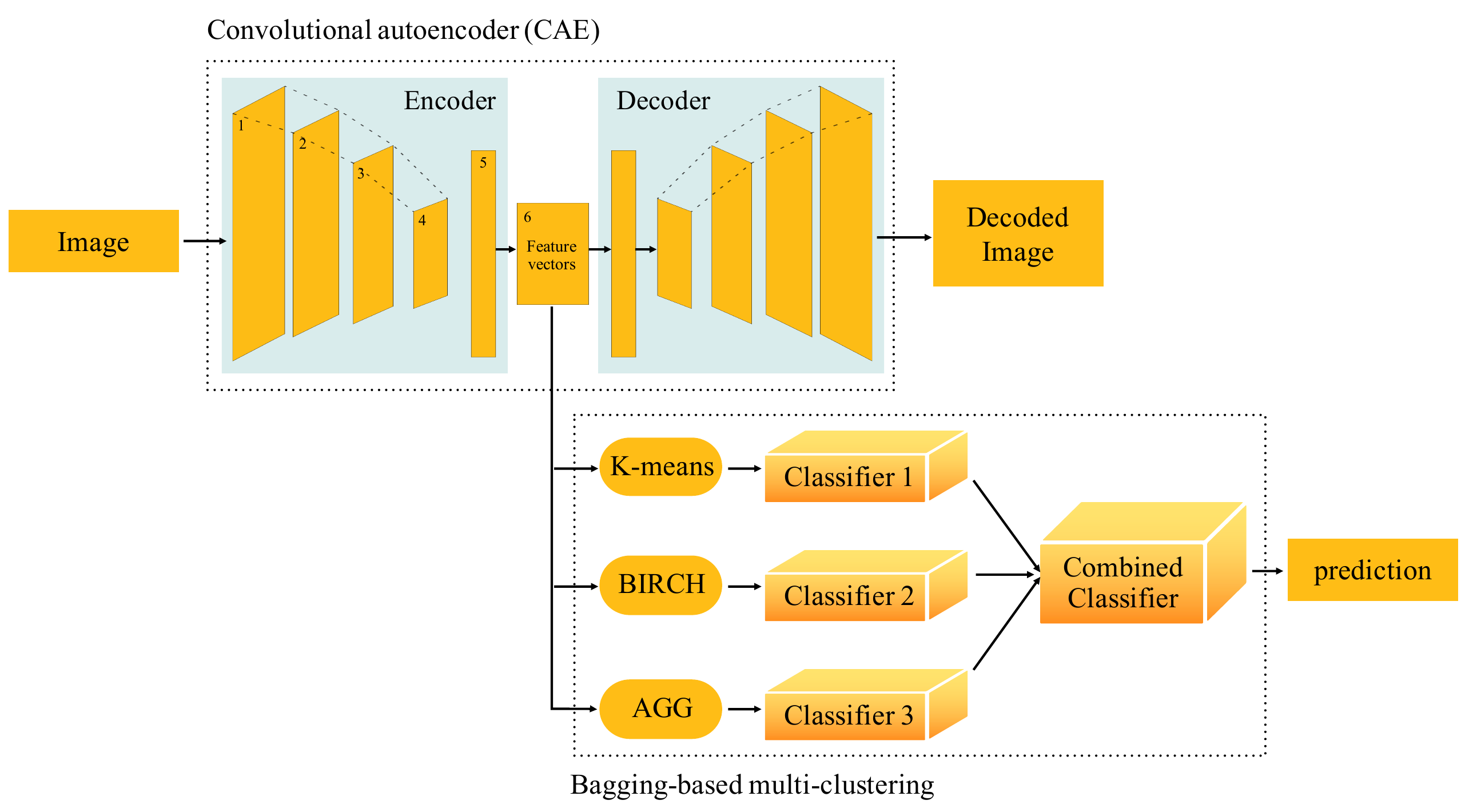}  % 图片路径
\caption{An illustration of the UML clustering process. We use the CAE to reduce the noise of the original image. Then the Bagging-based multi-clustering model is carried out on the galaxy according to the encoded data.}  % 图片标题
\label{fig2}    % 标签，用来引用
\end{figure*}
Following \cite{2022AJ....163...86Z}, we crop the original large-size image to a size of $28 \times 28$ and place the galaxy in the center of the image so that unnecessary noise interference is reduced. Then we use the convolutional autoencoders (CAE) algorithm to extract image information and compress dimensions through different convolution and pooling operations at each layer \citep{inproceedings,7434593}. CAE is an effective technique to extract image features. It can be used for automatic noise reduction without requiring any label information and can be
achieved by reconstructing the image \citep{inproceedings}. The detail of the CAE architecture is shown in Table~\ref{tbl-1cae}. The parameters and loss function of each layer of CAE that we adopt are the same as those used in \cite{2022AJ....163...86Z}. As seen from Figure \ref{fig:fig02}, after applying CAE for noise reduction, image features are effectively extracted, and image quality is significantly improved.
\begin{table}[htbp]%调节图片位置，h：浮动；t：顶部；b:底部；p：当前位置
	\centering
	\caption{The CAE Architecture \label{tbl-1cae}}
%	\label{tab:1}  
	\setlength{\tabcolsep}{4mm}
	\begin{tabular}{cccc}%表格中的数据居中，c的个数为表格的列数
		\hline\hline\noalign{\smallskip}	
		layer & type & stride & dimension  \\
		\noalign{\smallskip}\hline\noalign{\smallskip}
		  1& Convolution & .... & 28$\times$28$\times$128\\
		  2& Maxpooling & 2$\times$2  & 14$\times$14$\times$128\\
	    3& Convolution & ... &14$\times$14$\times$128\\
	    4& Maxpooling & 2$\times$2 & 7$\times$7$\times$128\\
		5& Unfolding  & ... & 6272\\
		  6& Full connection & ... & 40\\
		\noalign{\smallskip}\hline
	\end{tabular}
	%\tablecomments{We changed the size of the input image without changing the network structure}
\end{table}

\subsection{UML clustering process}

The process of the UML clustering is illustrated in Figure~\ref{fig2}. As demonstrated in \cite{2022AJ....163...86Z}, a single clustering model may be biased and return a misclustering result. Thus, we adopt the bagging-based multi-clustering method \citep{2022AJ....163...86Z} to give a more robust clustering result of our pre-processed $28\times28$ images after applying CAE to the images for noise reduction, As shown in Figure~\ref{fig2}, the same batch of data is inputted into three clustering models simultaneously (i.e., K-means, \citealt{10.2307/2346830}; AGG, \citealt{10.1093/comjnl/26.4.354,Murtagh2014}; and BIRTH, \citealt{10.1145/233269.233324}). Each model clusters the sample into 100 categories. Those categories derived by the three models are aligned by setting labels of K-means as the main one and matching the group that shows the highest frequency of the K-means label in the result of the other two models \citep[see Section 3.3 for details]{2022AJ....163...86Z}. Once the categories are aligned, We take the majority win-out strategy in voting, and those sources for which the three models reach no agreement in voting are discarded.

In visual classification, We randomly select a certain number of images (approximately 20 to 50) based on the number of galaxies in each group and display them on the same panel for visual classification. Three collaborators participated in this classification. We agree when two or more people have the same classification result for a specific galaxy. Otherwise, it is considered an unclassifiable galaxy. Thus, we divided them into five categories with physical meanings \citep[i.e., SPH, ETD, LTD, IRR, and UNC,][]{2022AJ....163...86Z}.
As a result, we finally obtain 8258 galaxies with reliable morphological labels, providing the basis for the following SML clustering process, and discard 9034 sources with inconsistent voting results in our UML clustering process.

\subsection{SML Clustering Process -- the GoogLeNet algorithm}
\begin{figure}[htbp]
\centering
\includegraphics[scale=0.9]{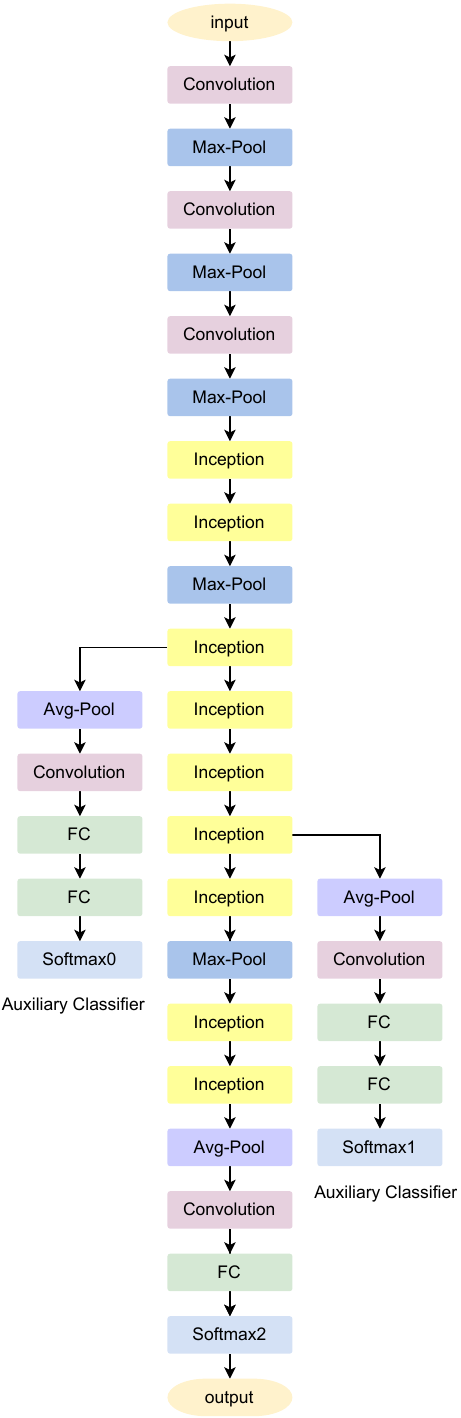}  % 图片路径
\caption{Flow chart of the GoogLeNet neural network structure used in this work. The tuple in the box represents the output shape of each layer. There are 22 layers, with no fully connected layers. The specific output parameters of each layer are given in Table ~\ref{tbl-1}.}  % 图片标题
\label{fig:fig04}    % 标签，用来引用
\end{figure}

To complete the classification of our sample, we take the 8528 UML well-classified sources as a training set and conduct SML to the rest 9034 galaxies.
As demonstrated by \cite{2023AJ....165...35F}, GoogLeNet performs well in the classification of deep-field galaxies in the classical neural network model. Therefore, we adopt GoogLeNet \citep{7298594} here as a supervised classification model.

\begin{table}[htbp]%调节图片位置，h：浮动；t：顶部；b:底部；p：当前位置
\begin{center}
	\caption{The GoogLeNet Architecture \label{tbl-1}}
%	\label{tab:1}  
	\setlength{\tabcolsep}{4mm}
	\begin{tabular}{cccc}%表格中的数据居中，c的个数为表格的列数
		\hline\hline\noalign{\smallskip}	
		type & stride & output size & depth \\
		\noalign{\smallskip}\hline\noalign{\smallskip}
		  Convolution & 7$\times$7/2 & 28$\times$28$\times$64 & 1 \\
		  Maxpooling & 3$\times$3/2 & 14$\times$14$\times$192 & 0 \\
		  Convolution & 3$\times$3/2 &14$\times$14$\times$192 & 0 \\
	    Maxpooling & 3$\times$3/2 &14$\times$14$\times$192 & 0\\
		  inception(3a) & ... & 14$\times$14$\times$256 & 2 \\
		  inception(3b) & ... & 14$\times$14$\times$480 & 2 \\
		  Maxpooling & 3$\times$3/2 &7$\times$7$\times$480 & 0\\
		  inception(4a) & ... & 7$\times$7$\times$512 & 2 \\
		  inception(4b) & ... & 7$\times$7$\times$512 & 2\\
		  inception(4c) & ... & 7$\times$7$\times$512 & 2\\
		  inception(4d) & ... & 7$\times$7$\times$528 & 2\\
		  inception(4e) & ... & 7$\times$7$\times$832 & 2\\
		  MaxPooling & 3$\times$3/2 &4$\times$4$\times$832 & 0\\
		  inception(5a) & ... & 4$\times$4$\times$832 & 2\\
		  inception(5b) & ... & 4$\times$4$\times$1024 & 2\\
		  AveragePooling & 7$\times$7/1 &1$\times$1$\times$1024 & 0\\
		  dropout(40 percent) & & 1$\times$1$\times$1024 & 0\\
		  linear & ... & 1$\times$1$\times$1000 & 1\\
		  softmax & ... & 1$\times$1$\times$1000 & 0\\
		\noalign{\smallskip}\hline
	\end{tabular}
 \end{center}
	\tablecomments{We change the size of the input image, but keep the network structure unchanged.}
\end{table}
\begin{table}[htbp]%调节图片位置，h：浮动；t：顶部；b:底部；p：当前位置
\begin{center}
	\caption{The number of sources in the training set and verification set \label{tbl-2}}
	\label{tab:2}  
	\setlength{\tabcolsep}{5mm}
	\begin{tabular}{ccc}%表格中的数据居中，c的个数为表格的列数
		\hline\hline\noalign{\smallskip}	
	galaxy type & Training set & verification set\\
		\noalign{\smallskip}\hline\noalign{\smallskip}
            SPH&1944&223\\
            ETD&650&65\\
            LTD&1091&136\\
            IRR&1342&143\\
            UNC&2385&279\\
            ALL&7412&846\\
		\noalign{\smallskip}\hline
	\end{tabular}
 \end{center}
\tablecomments{For data successfully classified by UML, the ratio of training set and verification set is 9:1.}
\end{table}

The structure of GoogLeNet is shown in Figure~\ref{fig:fig04}. The inception structure has two main contributions. One is to superposition more convolution in the case of the same size and extract more abundant features. The other is the simultaneous convolution re-aggregation on multiple sizes, which can extract features of different scales, making classification more accurate and improving efficiency. In the inception, the structure is to bring together features with solid correlations to accelerate convergence. The model parameters of each layer used in this work are described in Table \ref{tbl-1}.

In this part of the work, we use the 8258 well-classified galaxies obtained from the UML model as labeled data to classify the remaining 9034 galaxies. In order to avoid overfitting, we randomly divide the labeled data into the training (7412) and verification (846) sets with a fixed proportion of about 9:1 as shown in Table~\ref{tab:2} \citep{2023AJ....165...35F}. When training the GoogLeNet model, the algorithm's step size, learning rate, and depth are referred to \cite{2023AJ....165...35F}.

\section{Result and discussion}\label{sec:inspection}

\begin{table*}[htbp]%调节图片位置，h：浮动；t：顶部；b:底部；p：当前位置
	\centering
	\caption{Demographic of our result \label{tbl-5}}
%	\label{tab:5}
	\setlength{\tabcolsep}{5mm}
\begin{tabular}{ccccccc}%表格中的数据居中，c的个数为表格的列数
		\hline\hline\noalign{\smallskip}	
		Model & SPH & ETD & LTD & IRR & UNC & Totel \\
		\noalign{\smallskip}\hline\noalign{\smallskip}
		UML & 2664 & 1485 & 1227 & 715 & 2167 & 8258\\
		SML (i.e., GoogLeNet) & 2671 & 1647 & 1610 & 978 & 2128 & 9034 \\
		Totel & 5335 & 3132 & 2837 & 1693 & 4295 & 17292 \\
		\noalign{\smallskip}\hline
	\end{tabular}
	\tablecomments{Nearly a quarter of the images are classified as UNC due to the limitation of data quality, }
\end{table*}

 \begin{table*}[htbp]%调节图片位置，h：浮动；t：顶部；b:底部；p：当前位置
\begin{center}
	\caption{The fully classified catalog}
	\label{tab:6}  
	\begin{tabular}{ccccccccccc}%表格中的数据居中，c的个数为表格的列数
		\hline\hline\noalign{\smallskip}	
		Seq & R.A. & Dec. & $H_{\rm mag}$ & $z$ & $M_{\star}$ & $r_{e}$ & $n$ & $G$ & $M_{20}$ & Morphology \\
  ---&deg&deg&mag&---&$\log{M_{\odot}}$&kpc&---&---&---&---\\
  (1) & (2)&(3)&(4)&(5)&(6)&(7)&(8)&(9)&(10)&(11)\\
		\noalign{\smallskip}\hline\noalign{\smallskip}
1	&	150.55484 	&	1.98613 	&	21.44 	&	0.84 	&	10.17 	&	6.21 	&	2.01 	&	0.45 	&	-1.39 	&	4	\\
2	&	150.53548 	&	1.98611 	&	19.40 	&	0.58 	&	10.98 	&	3.10 	&	4.82 	&	0.56 	&	-1.87 	&	0	\\
3	&	150.50478 	&	1.98616 	&	22.95 	&	1.30 	&	10.22 	&	0.60 	&	0.20 	&	0.47 	&	-1.35 	&	1	\\
4	&	150.44777 	&	1.98596 	&	22.47 	&	1.45 	&	10.19 	&	3.32 	&	0.20 	&	0.50 	&	-0.79 	&	3	\\
5	&	150.48640 	&	1.98582 	&	21.38 	&	0.87 	&	10.11 	&	8.98 	&	0.90 	&	0.41 	&	-1.65 	&	4	\\
6	&	150.47160 	&	1.98549 	&	20.18 	&	0.72 	&	10.55 	&	1.84 	&	3.61 	&	0.56 	&	-1.81 	&	0	\\
7	&	150.49690 	&	1.98546 	&	22.68 	&	1.43 	&	10.42 	&	5.49 	&	0.27 	&	0.42 	&	-1.23 	&	4	\\
8	&	150.53648 	&	1.98500 	&	19.45 	&	0.58 	&	10.95 	&	3.75 	&	4.70 	&	0.58 	&	-1.85 	&	0	\\
9	&	150.48744 	&	1.98523 	&	22.17 	&	1.01 	&	10.06 	&	4.06 	&	1.01 	&	0.44 	&	-1.36 	&	3	\\
10	&	150.57190 	&	1.98463 	&	22.02 	&	1.34 	&	10.56 	&	4.37 	&	0.72 	&	0.46 	&	-1.33 	&	2	\\
11	&	150.57718 	&	1.98469 	&	22.88 	&	2.47 	&	10.11 	&	4.51 	&	0.24 	&	0.41 	&	-1.13 	&	3	\\
12	&	150.52214 	&	1.98479 	&	20.83 	&	1.18 	&	10.71 	&	2.60 	&	2.36 	&	0.50 	&	-1.68 	&	1	\\
13	&	150.39563 	&	1.98396 	&	20.56 	&	1.05 	&	10.63 	&	5.77 	&	2.25 	&	0.48 	&	-1.90 	&	2	\\
14	&	150.56590 	&	1.98427 	&	21.64 	&	0.98 	&	10.21 	&	6.80 	&	2.48 	&	0.45 	&	-1.52 	&	4	\\
15	&	150.55733 	&	1.98419 	&	20.98 	&	0.81 	&	10.17 	&	3.78 	&	5.70 	&	0.57 	&	-1.77 	&	0	\\
16	&	150.41916 	&	1.98263 	&	20.93 	&	2.08 	&	11.70 	&	2.09 	&	2.47 	&	0.53 	&	-1.72 	&	0	\\
17	&	150.48875 	&	1.98280 	&	20.82 	&	1.26 	&	10.80 	&	1.78 	&	1.53 	&	0.52 	&	-1.62 	&	0	\\
18	&	150.47710 	&	1.98277 	&	22.20 	&	1.27 	&	10.02 	&	3.61 	&	0.42 	&	0.44 	&	-0.86 	&	4	\\
19	&	150.48128 	&	1.98282 	&	23.19 	&	2.70 	&	10.09 	&	1.84 	&	0.49 	&	0.51 	&	-1.33 	&	4	\\
20	&	150.52061 	&	1.98173 	&	22.74 	&	1.91 	&	10.03 	&	2.63 	&	1.84 	&	\nodata	&	\nodata	&	4	\\
		\noalign{\smallskip}\hline
	\end{tabular}
 \end{center}
	\tablecomments{(1) Sequential number identifier; (2) Right ascension expressed in decimal degrees; (3) Declination expressed in decimal degrees; (4) Magnitude in H band; (5) redshift; (6)Stellar mass; (7) effective radius; (8) S\'{e}rsic index; (9) Gini coefficient; (10) the normalized second-order moment of the brightest 20\% of the galaxy flux; (11) Morphology type: 0,1,2,3 and 4 represent SPH, ETD, LTD, IRR, and UNC respectively. (The full table is available online in machine-readable form.)}
\end{table*}

\subsection{Overall morphological classification results}

By combining UML with SML (i.e., the GoogLeNet model), we derive a complete morphology classification result of 17292 galaxies selected in the COSMOS-DASH field (Table~\ref{tbl-5}), which includes 5335 SPHs, 3132 ETDs, 2837 LTDs, 1693 IRRs, and 4295 UNCs. Part of the result is shown in Table~\ref{tab:6}.

\begin{figure*}[htbp!]
\centering
\includegraphics[scale=0.5]{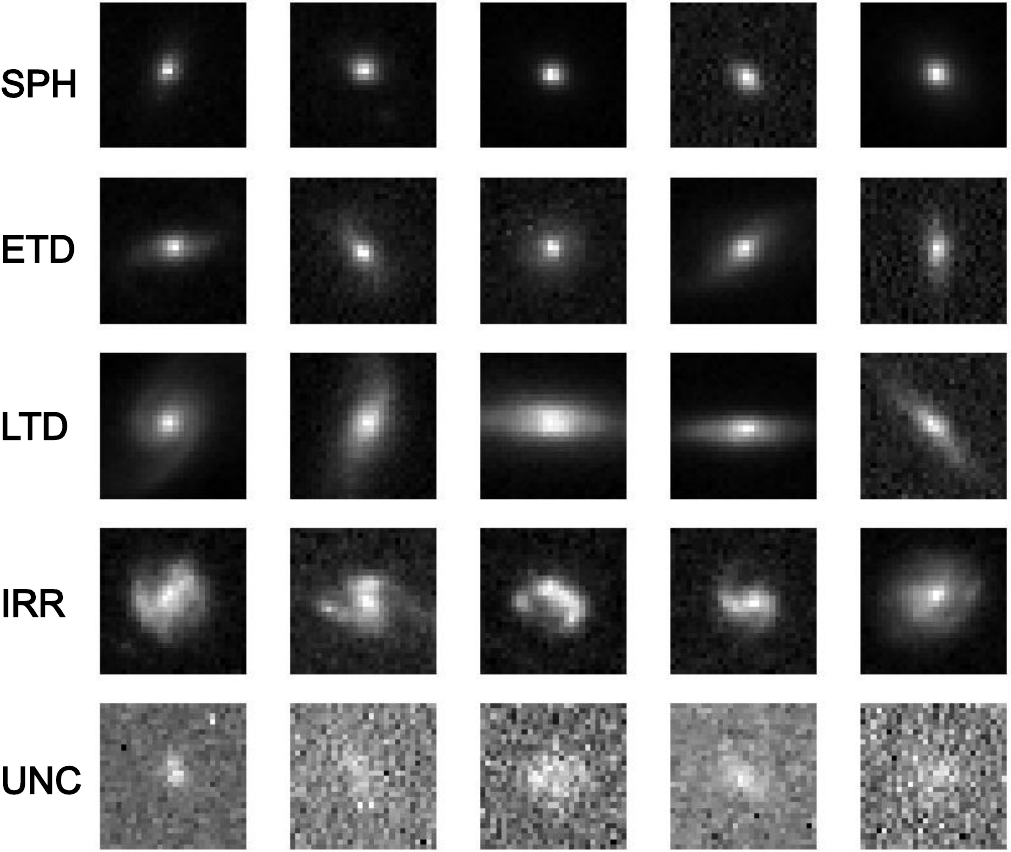} 
\caption{Examples of galaxies that finally divided into five categories.}  % 图片标题
\label{fig:fig03}    % 标签，用来引用
\end{figure*}

We sample the complete results of the classification for inspection and find that among which SPH is the most concentrated and brightest; ETD is slightly dim, with a bright nucleus in the center and a relatively concentrated luminosity. Most LTDs have an obvious nuclear sphere and spiral arm, while the luminosity is more diffuse. IRR does not have an apparent regular shape but can be identified as a galaxy. UNC is mainly shown in pictures with a meager signal-to-noise ratio, and it is impossible to identify whether there is a galaxy or what kind of galaxy it is. Figure \ref{fig:fig03} shows the pictures of randomly selected galaxies for each morphological type in the obtained label samples from the UML method. It can be seen from the pictures that the morphology types are distinguishable.

\begin{figure*}[htbp!]
\centering
\subfigure{
\includegraphics[width=7.8cm]{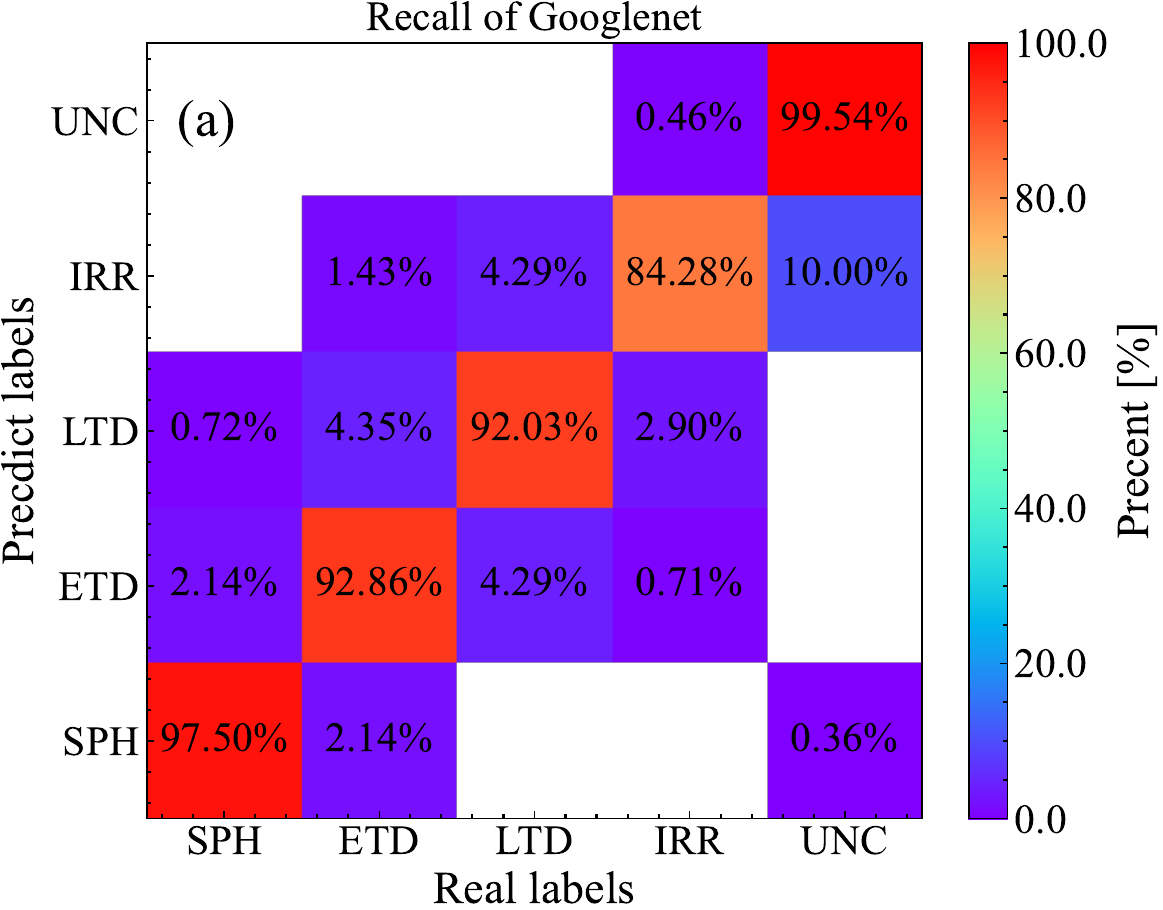}
}
\quad
\subfigure{
\includegraphics[width=7.8cm]{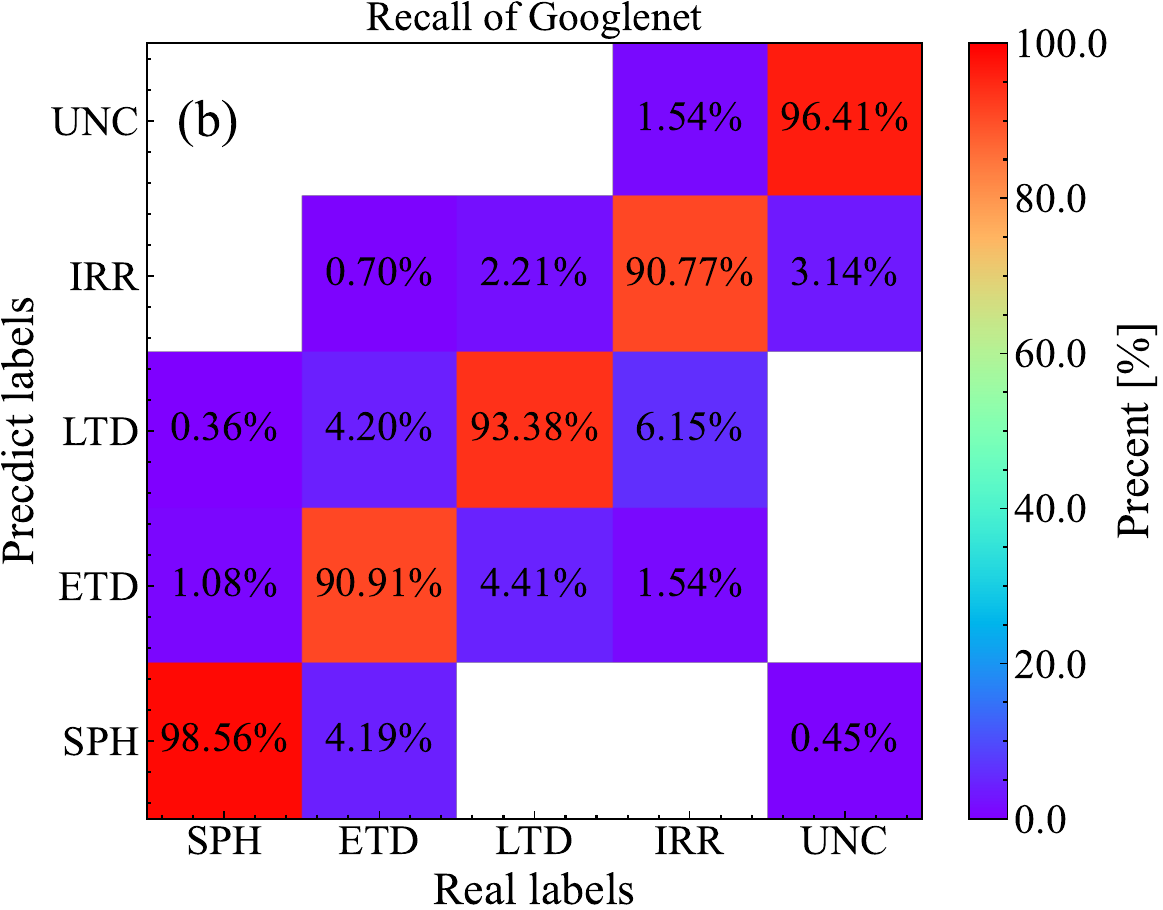}
}
\caption{The precision (panel a) and recall (panel b) of the GoogLeNet model. The high precision and recall rates indicate that GoogLeNet shows a good performance in classifying galaxies of all types, with a low probability that the various classes of galaxies are confused.}
\label{fig9} 
\end{figure*}

In Table~\ref{tbl-4}, we present the classification accuracies of the GoogLeNet model, which are larger than 90\% for all five types. Also, we test
the distribution of the verification set and training set in the physical parameter space, and the verification set can cover the entire parameter space well. The precision and recall of Figure \ref{fig9} are based on verification set estimation. The average precision and recall are both over 90\%, indicating that GoogLeNet has a good performance in classifying images of galaxies \citep{2023AJ....165...35F}, with a low probability that the various classes of galaxies are confused. Among them, the recognition accuracies of SPH and UNC are higher than those of other classes. We conclude from our analysis that SPH and UNC have more distinct features and therefore have better training in the model. It is typical for SPH to be misclassified as ETD because both galaxy populations exhibit smooth contours, and there is no strict boundary between them. Some LTDs have distinct nuclear sphere structures but not distinct spin arms, leading to misclassification as IRRs.

\begin{table*}[htbp]%调节图片位置，h：浮动；t：顶部；b:底部；p：当前位置
\begin{center}
	\caption{The accuracy of GoogLeNet model\label{tbl-4}}
%	\label{tab:4} 
	\setlength{\tabcolsep}{7mm}
	\begin{tabular}{ccc}%表格中的数据居中，c的个数为表格的列数
		\hline\hline\noalign{\smallskip}	
		Galaxy type & Accuracy of training set & Accuracy of the verification set  \\
		\noalign{\smallskip}\hline\noalign{\smallskip}
		SPH & 100.0\% & 97.84\% \\
		ETD & 100.0\% & 90.90\% \\
		LTD & 100.0\% & 93.38\% \\
		IRR & 100.0\% & 90.76\% \\
		UNC & 100.0\% & 96.41\% \\
		\noalign{\smallskip}\hline
	\end{tabular}
 \end{center}
	\tablecomments{Our method shows a good classification accuracy in all types of galaxies, which also indicates the robustness of results from our UML method.}
\end{table*}

\subsection{t-SNE test}

\begin{figure*}[htbp!]
\centering
\subfigure{
    \begin{minipage}[]{.3\linewidth}
        \centering
        \includegraphics[scale=0.45]{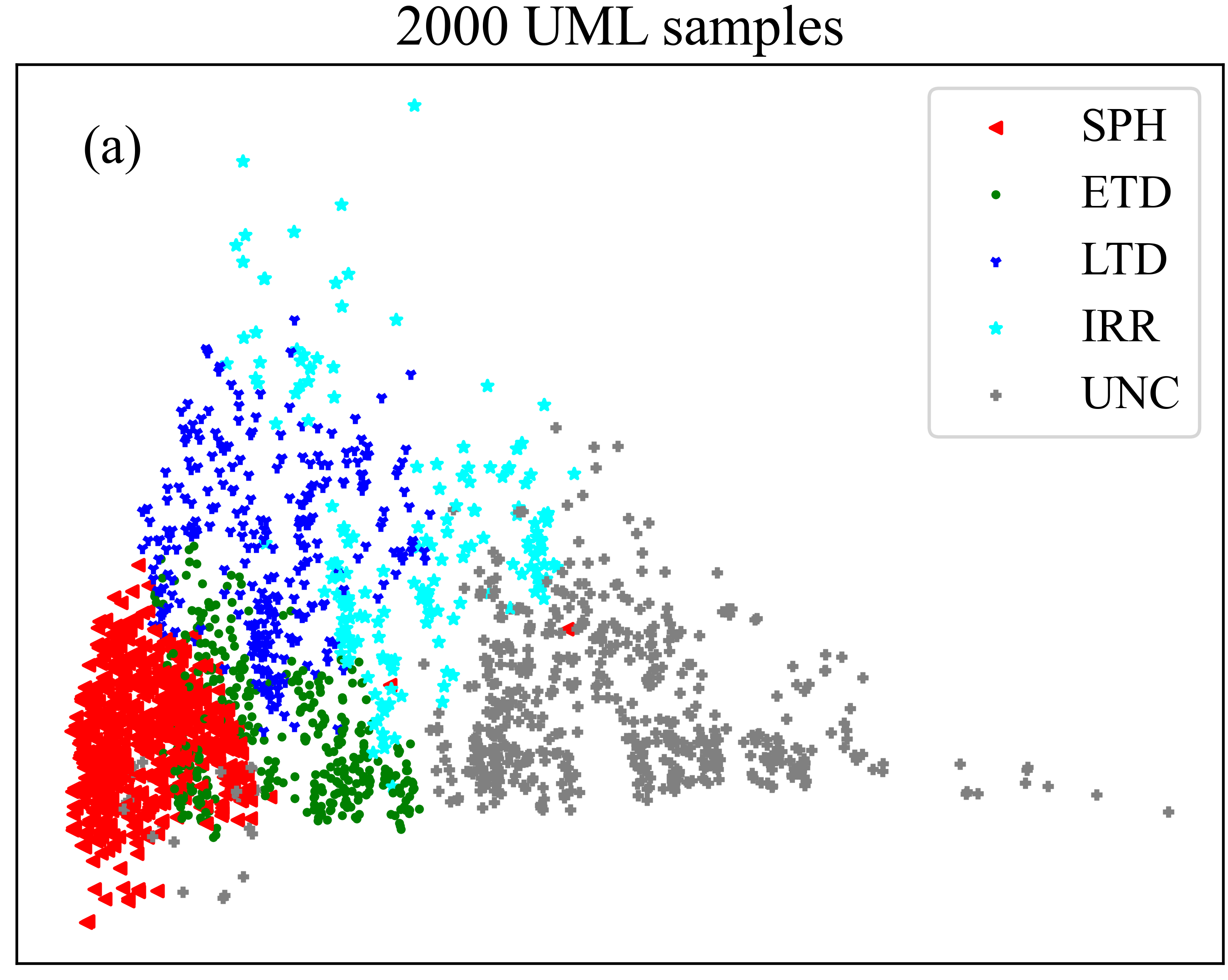}
    \end{minipage}
    }
\subfigure{
 	\begin{minipage}[]{.3\linewidth}
        \centering
        \includegraphics[scale=0.45]{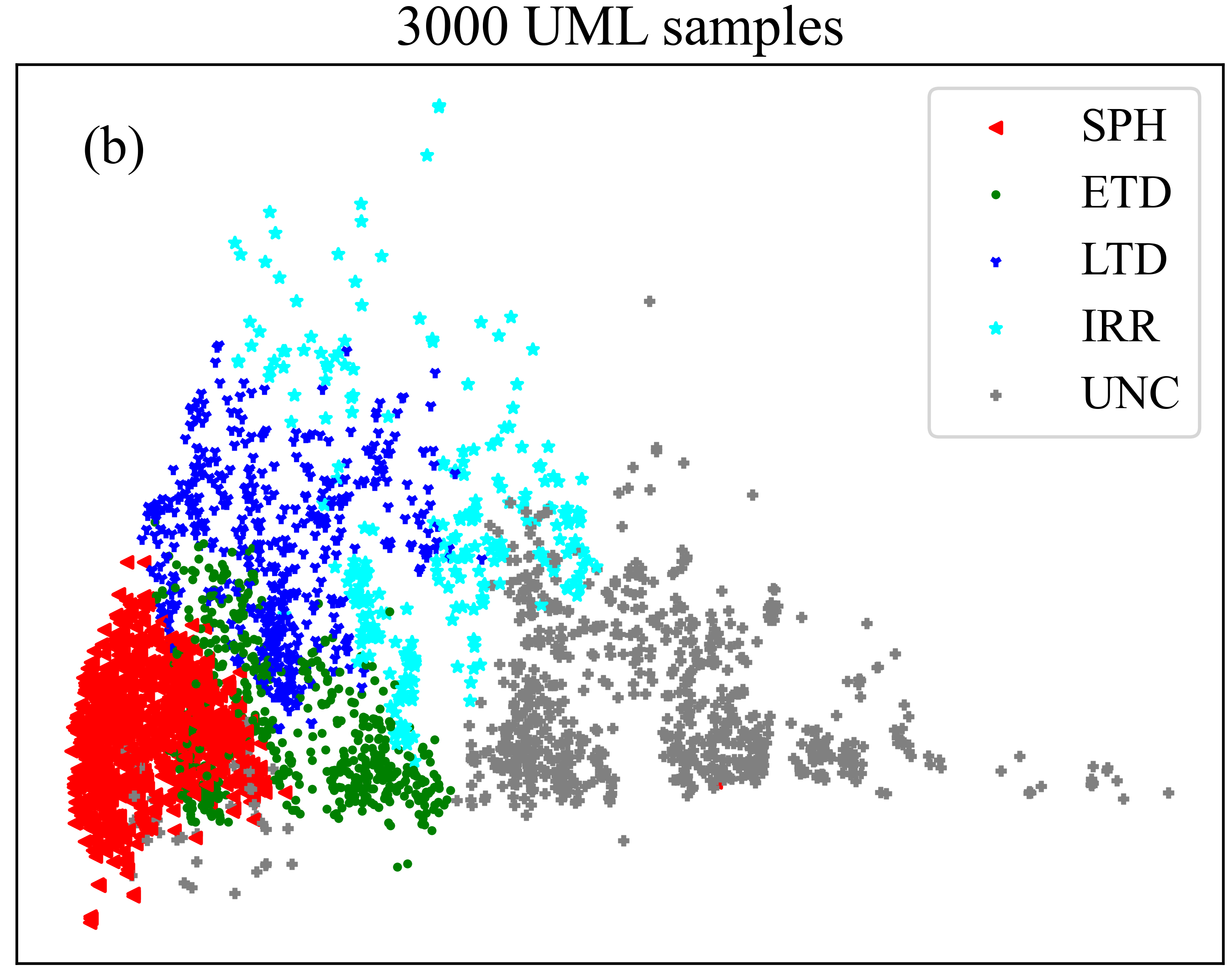}
    \end{minipage}
}
\subfigure{
 	\begin{minipage}[]{.3\linewidth}
        \centering
        \includegraphics[scale=0.45]{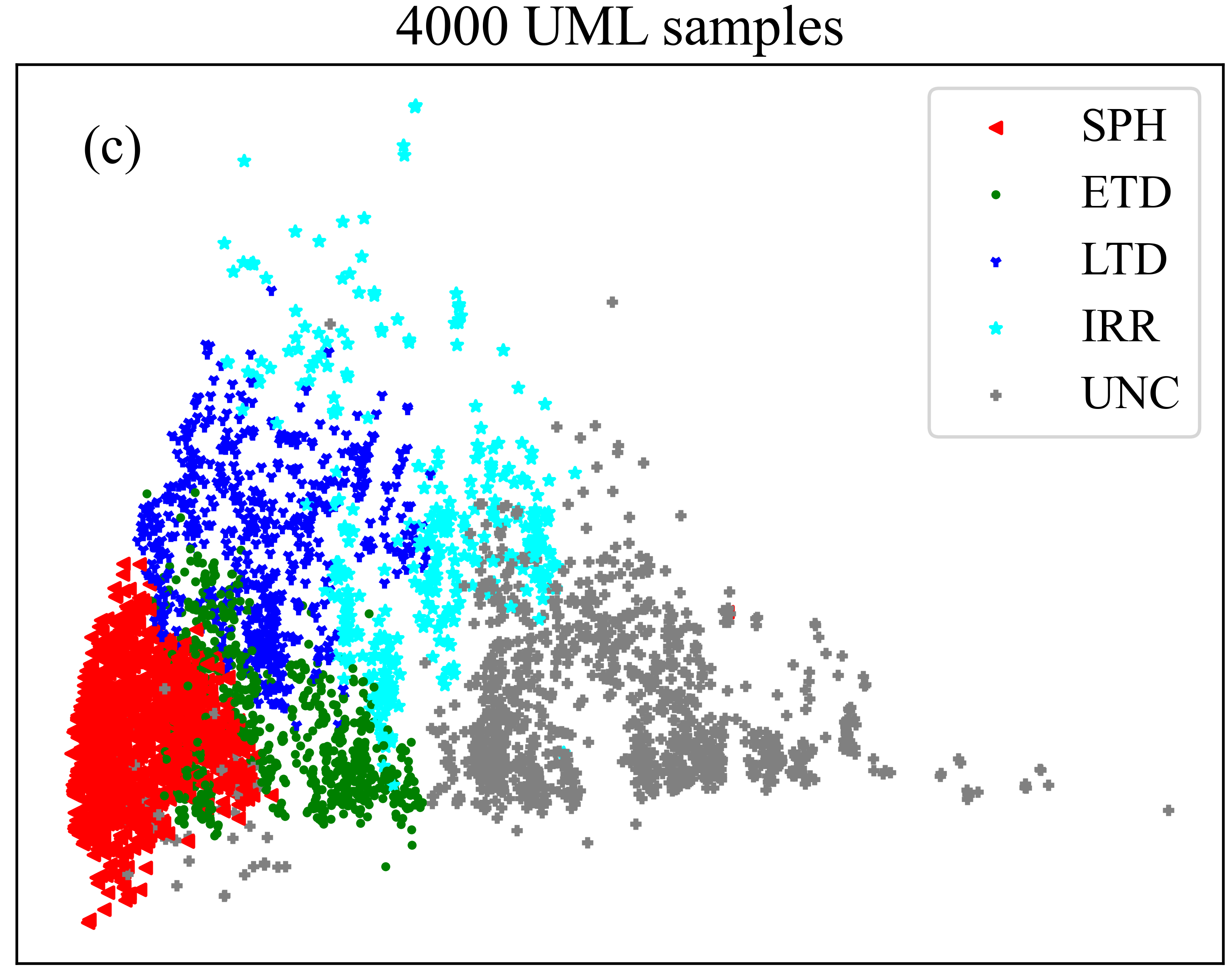}
    \end{minipage}
}
\subfigure{
    \begin{minipage}[]{.3\linewidth}
        \centering
        \includegraphics[scale=0.45]{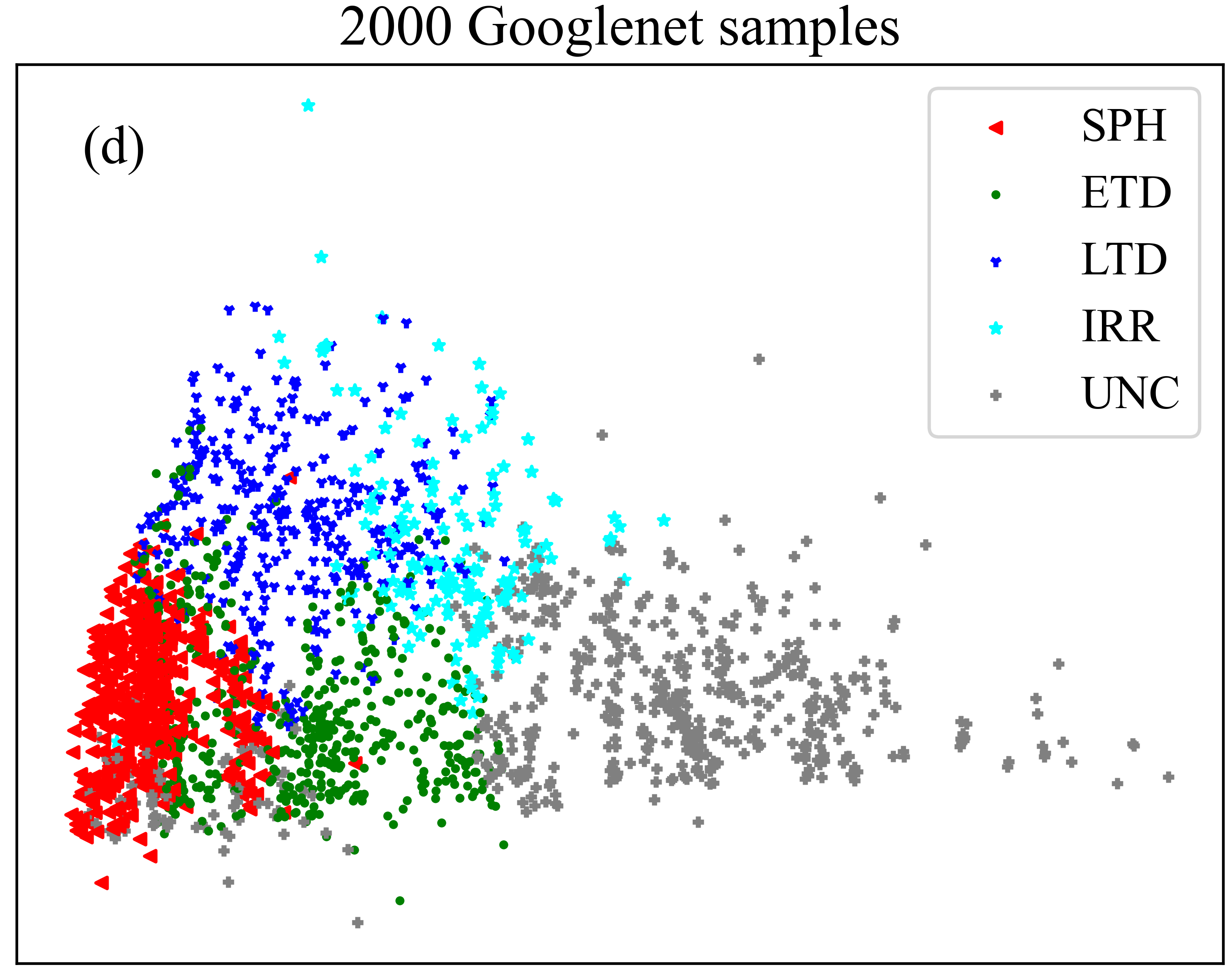}
    \end{minipage}
}
\subfigure{
 	\begin{minipage}[]{.3\linewidth}
        \centering
        \includegraphics[scale=0.45]{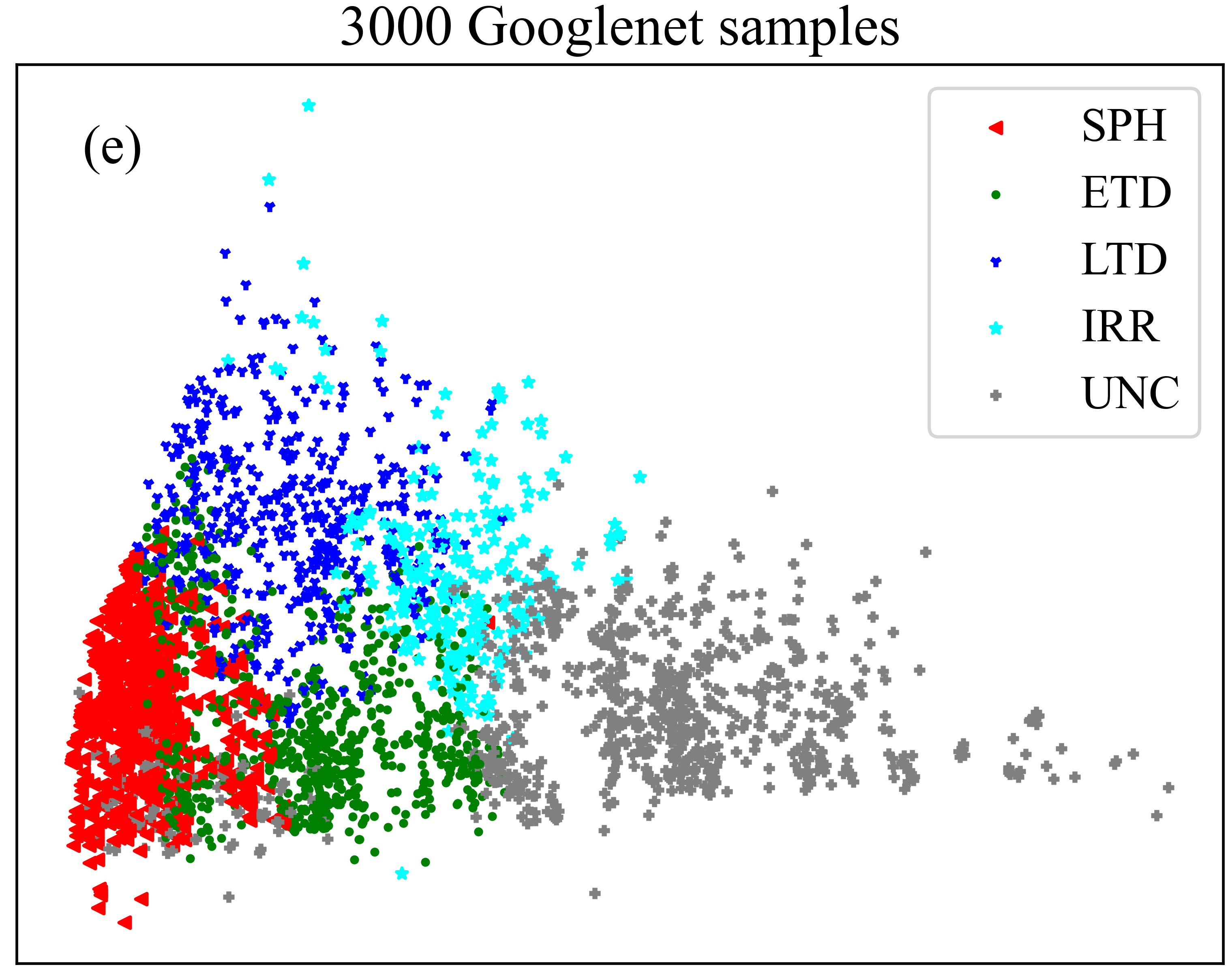}
    \end{minipage}
}
\subfigure{
 	\begin{minipage}[]{.3\linewidth}
        \centering
        \includegraphics[scale=0.45]{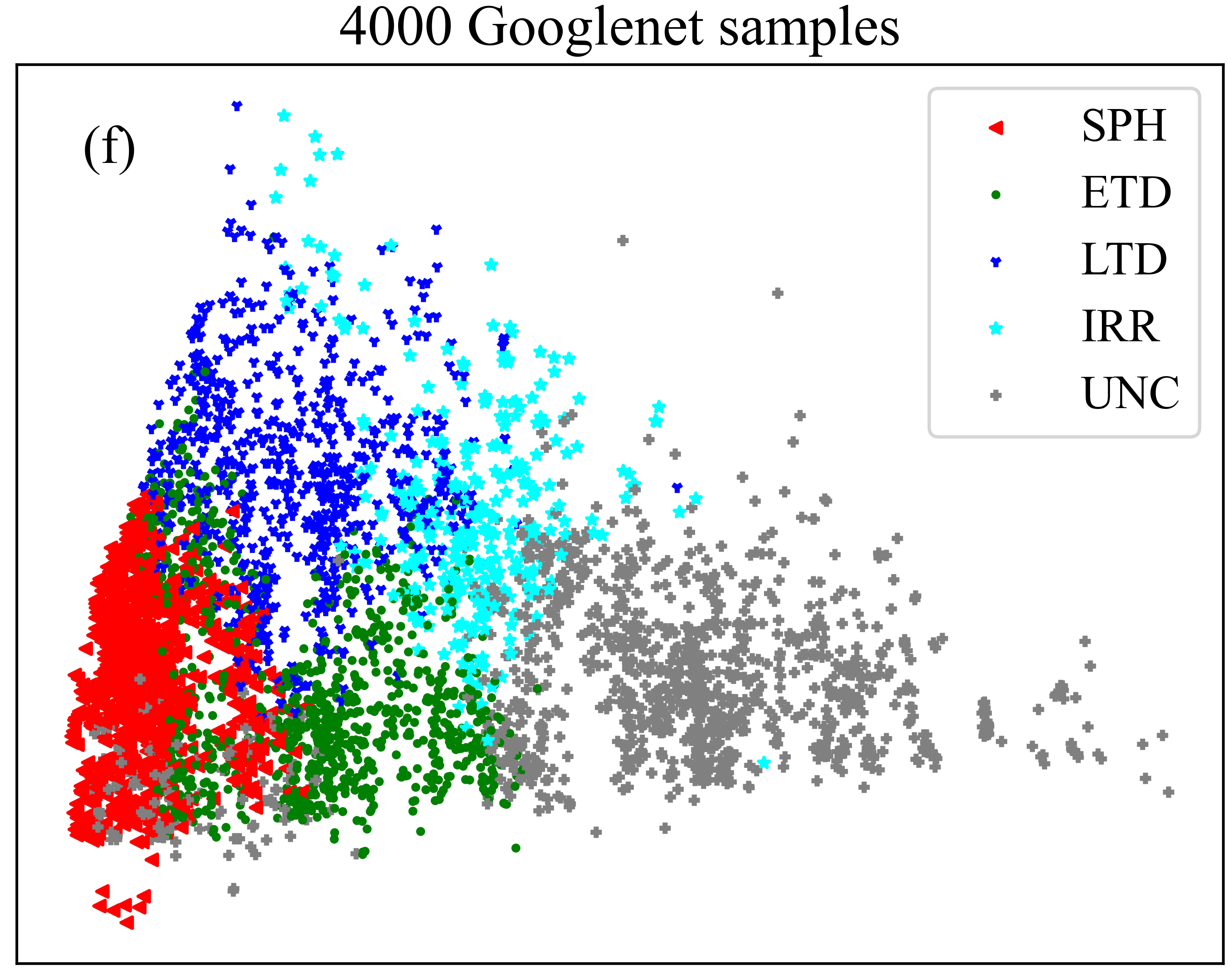}
    \end{minipage}
}

\subfigure{
    \begin{minipage}[]{.3\linewidth}
        \centering
        \includegraphics[scale=0.45]{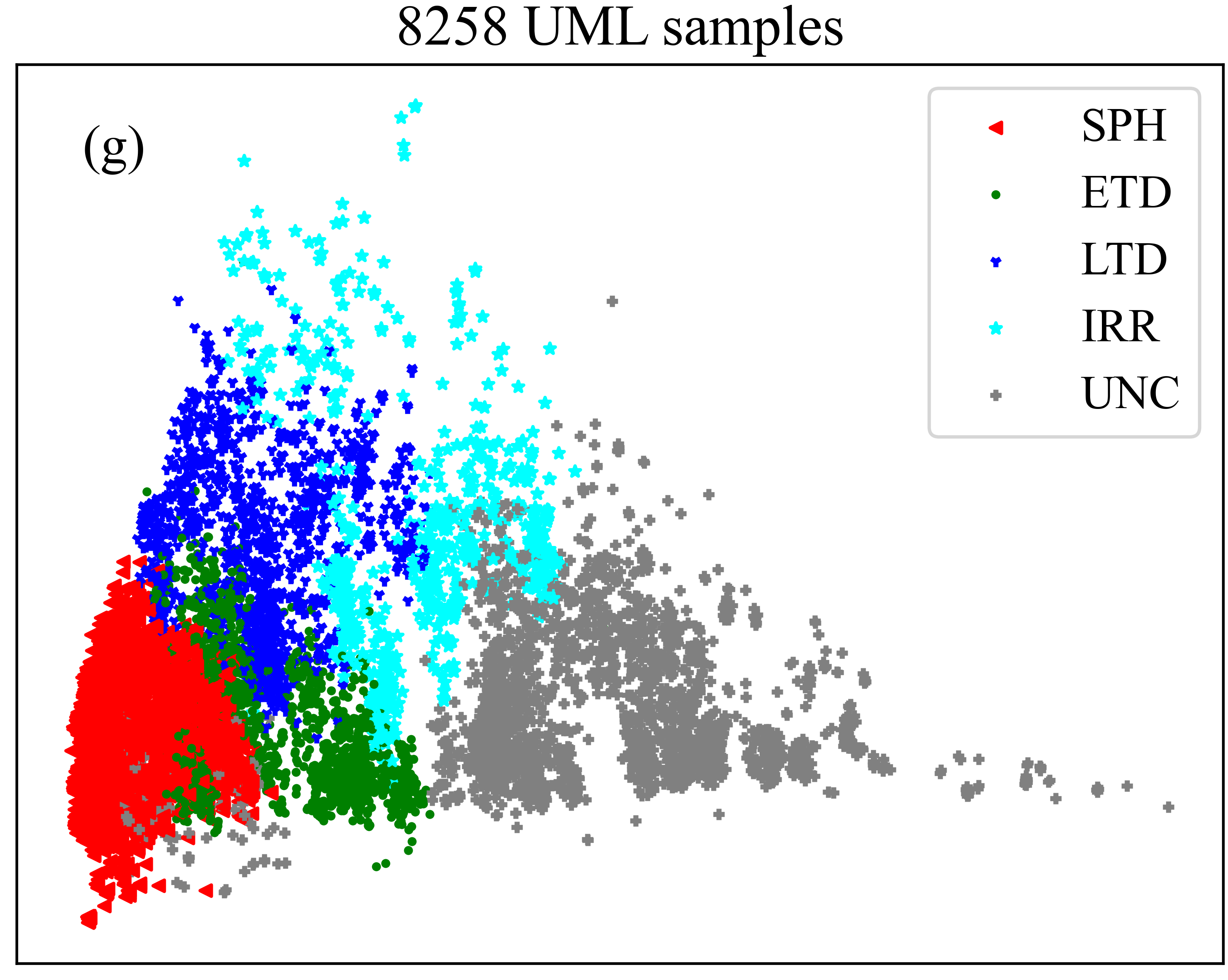}
    \end{minipage}
}
\subfigure{
 	\begin{minipage}[]{.3\linewidth}
        \centering
        \includegraphics[scale=0.45]{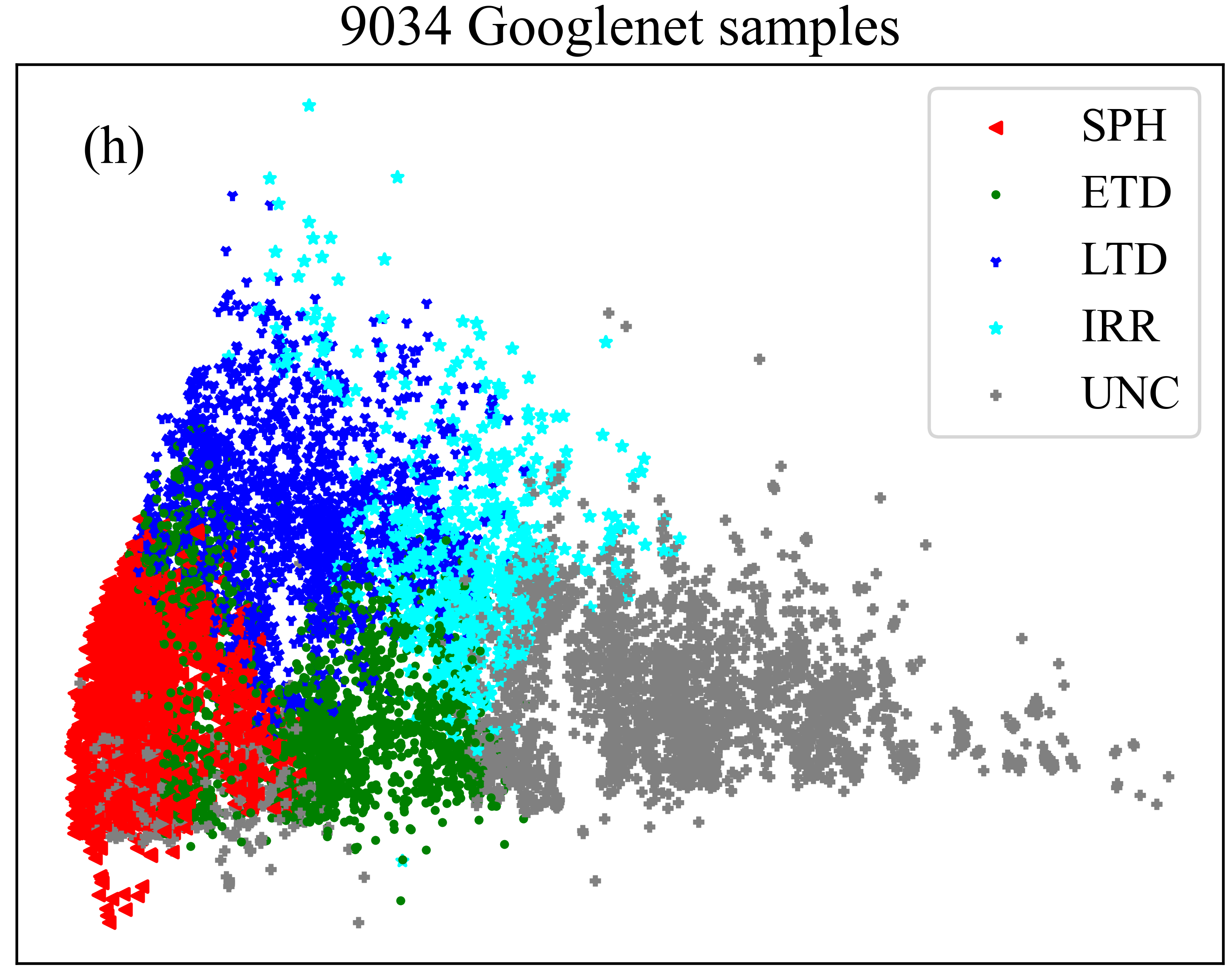}
    \end{minipage}
}
\subfigure{
 	\begin{minipage}[]{.3\linewidth}
        \centering 
        \includegraphics[scale=0.45]{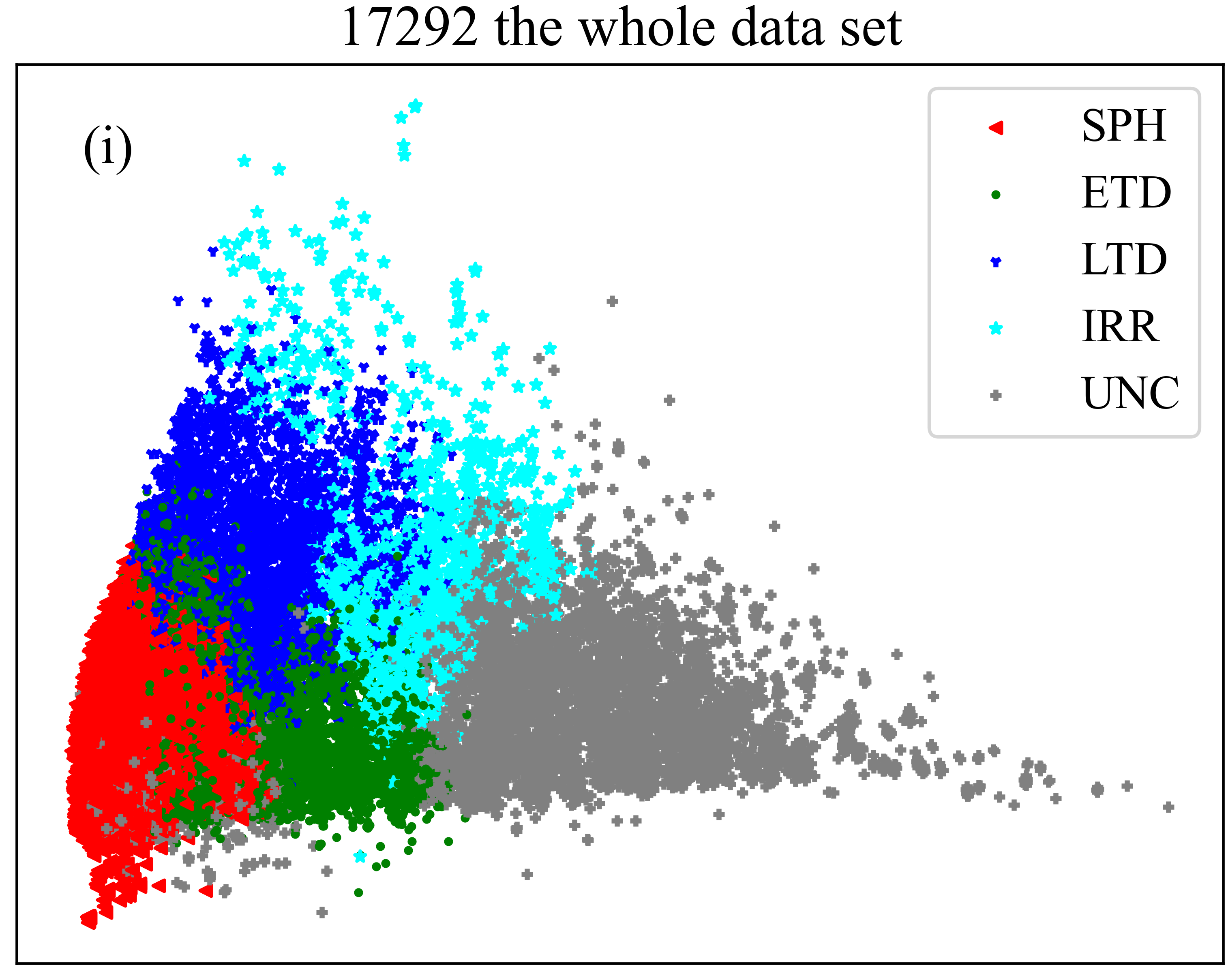}
    \end{minipage}
}
\caption{Panel (a), (b), and (c) are the t-SNE diagrams of randomly selected 2000, 3000, and 4000 galaxies that are labeled by the UML method, respectively. Panel (d), (e), and (f) are the t-SNE diagrams of 2000, 3000, and 4000 galaxies that are labeled by SML (i.e., GoogLeNet). Panel (g) and (h) are the t-SNE diagrams of the subsamples labeled by the UML and SML (i.e., GoogLeNet) methods, respectively. Panel (i) shows the t-SNE diagram of the full sample. Different types of galaxies are clustered in different regions within the two-dimensional parameter space. As the galaxy number increases, the five categories of galaxies show a clear trend of clustering. The categories with similar features are clustered together with a small overlap at the edges, which is due to the partial similarity of galaxy morphology during their evolutionary history. Our method shows a good performance in clustering galaxies.}
\label{fig:fig05}
\end{figure*}

The t-SNE graph is an efficient way to map high-dimensional data to a low-dimensional space and to transform the clustering results into dimensions suitable for inspection \citep{JMLR:v9:vandermaaten08a}. We sample the results of the five classes of galaxies that were finally classified by the UML and GoogLeNet model for 2000, 3000, and 4000 times using the t-SNE technique. As shown in Figure~\ref{fig:fig05}, the five categories of galaxies show a clear trend of clustering on the screen as the sampling time increases. Galaxies with similar features are clustered together. In each category, there is a small amount of overlay at the edges of the populations, which is caused by morphological similarities and is expected by galaxy morphological evolution.

The following conclusions can be drawn from Figure~\ref{fig:fig05}: (1) The UML method provides a feasible prior sample, which is reflected in the t-SNE graph that the distributions of all galaxy types tend to be stable. (2) The GoogLeNet model trained by the result of UML successfully classified the remaining sources, and keep the aggregation degree consistent with the UML method. There are apparently distinguishable boundaries for all types of galaxies, indicating the reliability of our classification method.

\subsection{Test of Morphological Parameters}

Galaxy morphology parameters play an essential role in the description of the physical properties of galaxies. Different categories of galaxies show different physical properties, and the correspondence between the visual classification results and the physical properties of massive galaxies can effectively reflect the reliability of our result \citep{10.1111/j.1365-2966.2007.12627.x,2018ApJ...855...10G,2022AJ....163...86Z}. In this section, we analyze the classification results using galaxy morphology parameters. Since most of the UNC images have a meager signal-to-noise ratio, morphological parameters are difficult to measure and might have large uncertainties. On the other hand, ignoring the UNC sources would not affect the analysis of other classes, so we do not discuss the nature of UNC in this section.

\subsubsection{Parametric Measurements}

To derive the galaxy morphology parameters, we used the GALFIT package \citep{pengDetailedStructuralDecomposition2002} and GALAPAGOS software \citep{bardenGalapagosPixelsParameters2012} to fit galaxy surface brightness profiles with a single S\'{e}rsic model and measure the S\'{e}rsic index $n$ and the effective radius $r_e$ for each galaxy.

\begin{figure*}[htbp]
\centering
\includegraphics[scale=0.4]{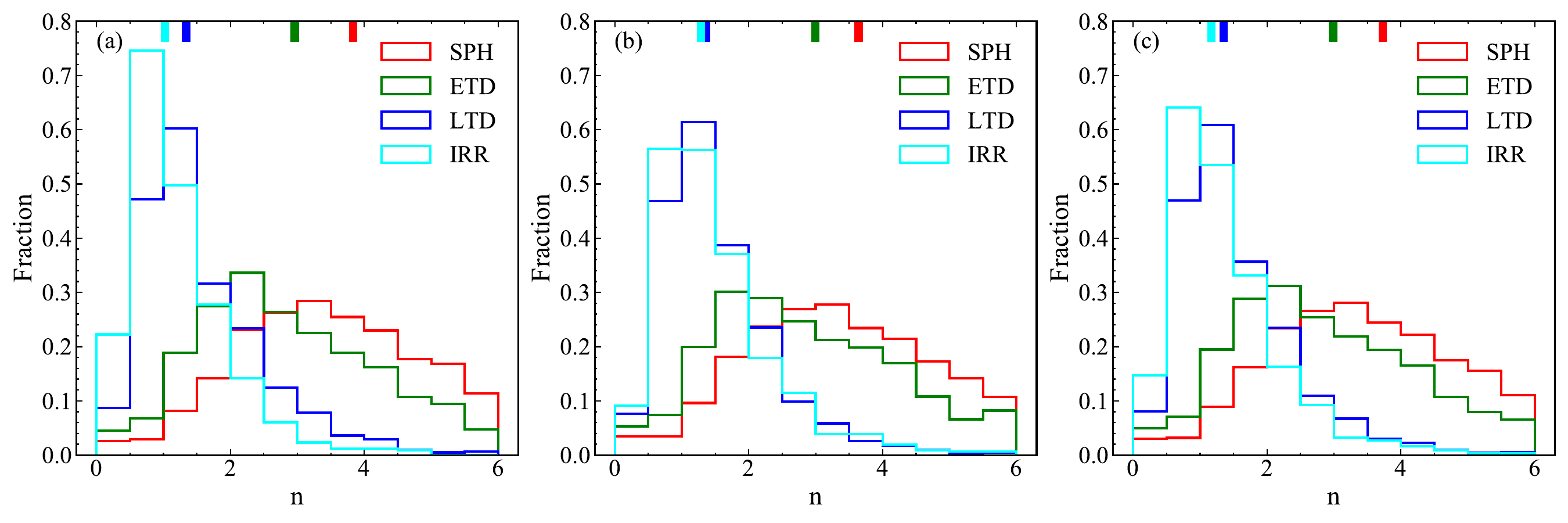}  % 图片路径
\caption{Distributions of the S\'{e}rsic index of galaxies (red: SPH, green: ETD, blue: LTD, and cyan: IRR). Panels (a), (b), and (c) represent the distributions of the UML clustering results, the GoogLeNet classification results, and the overall classification results, respectively. The bars at the top of each panel represent the median value of each class. As shown in panel (a), among the 8258 galaxies clustered by the UML method, the median S\'{e}rsic indexes of IRR, LTD, ETD, and SPH are 1.03, 1.34, 2.96, and 3.83, respectively, which show a gradually increasing trend from IRR to SPH. The classification results of GoogLeNet (panel b) and the overall classification results (panel c) maintain the same trend, which is consistent with the characteristic of the various types of galaxies.}  % 图片标题
\label{fig:fig06}    % 标签，用来引用
\end{figure*}

The distributions of the S\'{e}rsic index are shown in Figure~\ref{fig:fig06}. Panel (a) shows that among the 8258 galaxies successfully clustered by the UML method, the median S\'{e}rsic index of IRR, LTD, ETD, and SPH are 1.03, 1.34, 2.96, and 3.83, respectively, with a gradually increasing trend. In panel (b), the median S\'{e}rsic index of IRR, LTD, ETD, and SPH were 1.29, 1.36, 3.00, and 3.64, respectively; in panel (c), the median S\'{e}rsic indexes of IRR, LTD, ETD, and SPH are 1.17, 1.35, 2.98 and 3.73, respectively. The classification results of GoogLeNet (panel b) and the overall classification results (panel c) both share similar distributions for the four galaxy types and the same increasing trend from IRR to SPH with the UML sample, which is consistent with the expected correlation between this parameter and galaxy morphology.
% the physical properties of the various types of galaxies on this parameter.

\begin{figure*}[htbp]
\centering
\includegraphics[scale=0.4]{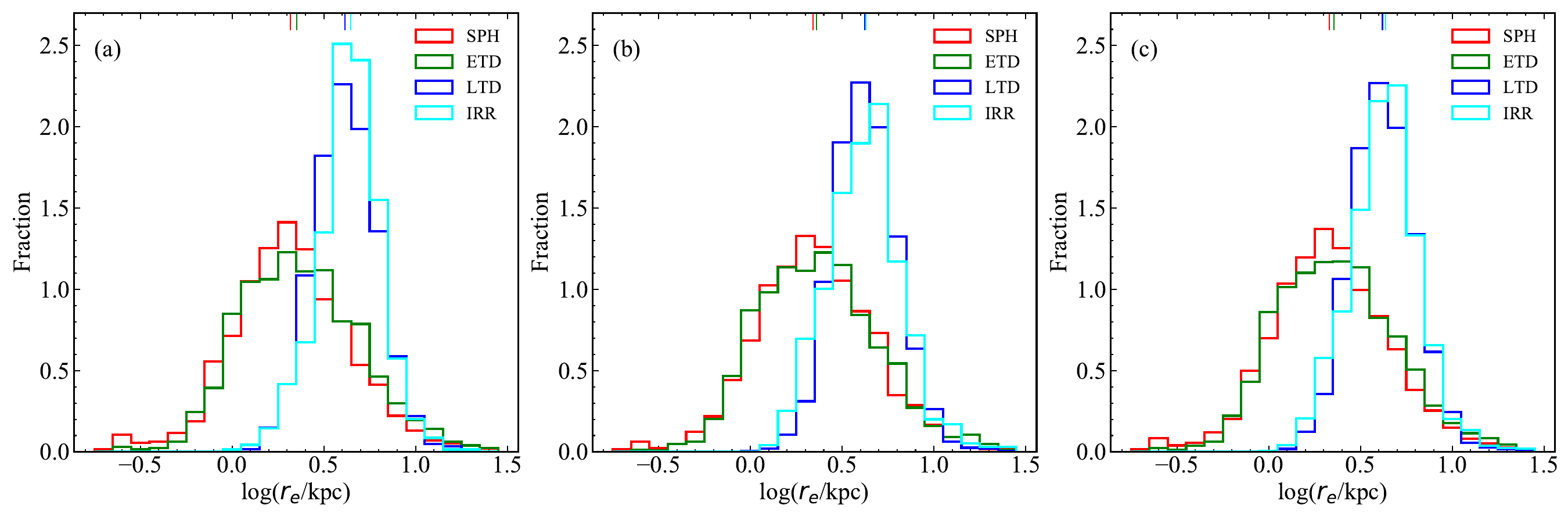}  % 图片路径
\caption{Distributions of the effective radii of galaxies. The sequence and symbols are the same as Figure~\ref{fig:fig06}. The median value of the effective radii increases from SPH, ETD, LTD to IRR. The distribution of the effective radius of different classes of galaxies is consistent with the morphological evolution history of galaxies.}  % 图片标题
\label{fig:fig07}    % 标签，用来引用
\end{figure*}

The effective radius distributions of the four classes are shown in Figure~\ref{fig:fig07}. Among the 8258 galaxies clustered by the UML method (panel a), the median effective radii of the four classes (i.e., SPH, ETD, LTD, and IRR) are 2.09, 2.24, 4.07, and 4.47 kpc, respectively. Among the 9034 galaxies classified by the GoogLeNet model (panel b), the median effective radii of SPH, ETD, LTD, and IRR are 2.19, 2.29, 4.17, and 4.27 kpc. In the total sample of 17,292 galaxies (panel c), the median effective radii of SPH, ETD, LTD, and IRR are 2.14, 2.29, 4.17, and 4.37 kpc, respectively. The median distribution of effective radii of galaxies increases from SPH, ETD, LTD to IRR.

In short, the distributions of the S\'{e}rsic index and effective radius of different classes of galaxies derived from our method are consistent with the expected correlations between galaxy morphologies and these structure parameters.

\subsubsection{Nonparametric Measurements}

Using the Morpheus program \citep{2007ApJ...669..184A}, we calculate the nonparametric morphological parameters Gini coefficient ($G$) and the normalized second-order moment of the brightest 20\% of the galaxy's flux ($M_{20}$) for all galaxies in our sample. Thus, we can investigate the correspondence between the galaxy morphological classification results and the physical relations between the various types of galaxies.

The Gini coefficient ($G$) indicates the flux distribution of galaxies \citep{abrahamNewApproachGalaxy2003}. Following \cite{lotzNewNonparametricApproach2004}, it can be calculated as:
\begin{equation}
    G=\frac{1}{\overline{f}n(n-1)}\sum_{i=0}^n(2i-n-1)f_i,
\end{equation}
where n is the number of pixels of the galaxy, $f_i$ is the pixel flux value sorted in ascending order, and $\overline{f}$ represents the mean over the pixel values. $M_{20}$ is the normalized second-order moment of the brightest 20\% pixels of the galaxy defined as:
\begin{equation}
    M_{tot}=\sum_{i}^nM_i=\sum_i^nf_i[(x_i-x_c)^2+(y_i-y_c)^2]
\end{equation}
\begin{equation}
    M_{20}=\log_{10}\frac{\sum_iM_i}{M_{tot}},\ \mathrm{while}\ \sum_if_i<0.2f_{tot},
\end{equation}
where $f_{tot}$ is the total flux of the galaxy, $f_i$ is the flux value of each pixel i, $(x_i,\ y_i)$ is the position of pixel i, and $(x_c,\ y_c)$ is the center of the image. \cite{lotzNewNonparametricApproach2004} developed $M_{20}$ to trace the spatial distribution of bright nuclei, bars, and off-center clusters. The $G$--$M_{20}$ diagram is often used to test the separation of different classes of galaxies (e.g., \cite{lotzEvolutionGalaxyMergers2008}; \cite{rodriguez-gomezOpticalMorphologiesGalaxies2019}).

\begin{figure}[htbp]
\centering
\includegraphics[scale=0.45]{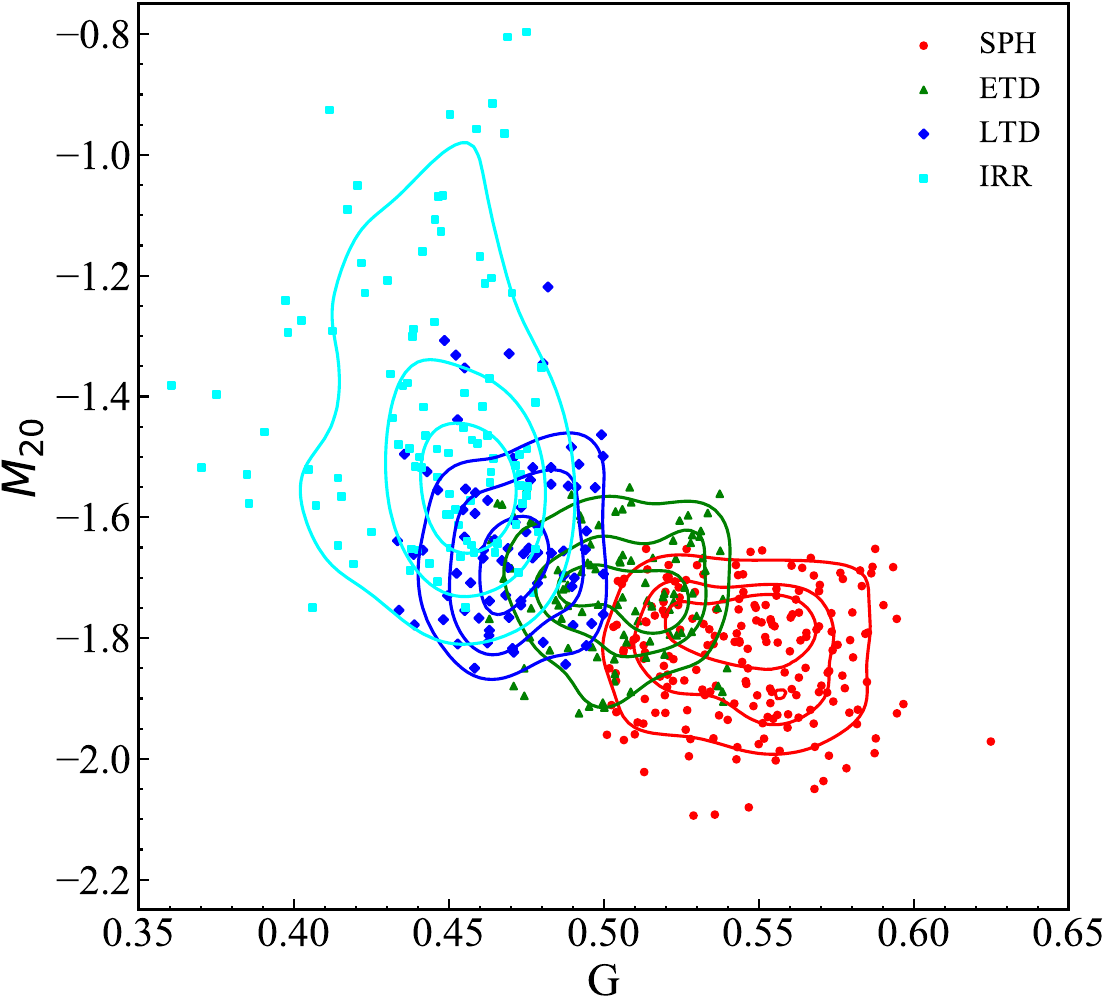}  % 图片路径
\caption{Distributions of galaxies in the $G$--$M_{20}$ parameter space (red: SPH, green: ETD, blue: LTD, and cyan: IRR). The contour levels indicate 20\%, 50\%, and 80\% of the corresponding classes from the inside to the outside. Individual data points are randomly selected from the four classes. $M_{20}$ decreases with the trend of IRR, LTD, ETD, and SPH, while $G$ increases from IRR to SPH. The galaxy classes are distinguishable in this diagram.}  % 图片标题
\label{fig:fig08}    % 标签，用来引用
\end{figure}

We plot the distribution of the four types of galaxies in the $G$--$M_{20}$ space. As shown in Figure ~\ref{fig:fig08}, various types of galaxies are well distinguished in the $G$--$M_{20}$ space. The Gini coefficient of galaxies gradually increases from IRR to SPH, while the value of $M_{20}$ slowly decreases. SPH galaxies tend to have the largest Gini coefficient and the smallest $M_{20}$. The overall trend from IRR to SPH in this diagram is in good agreement with the expected variations between these four morphology types, which further suggests the robustness of our two-step method to morphologically classify galaxies.

\section{conclusion} \label{sec:results}
In this paper, we apply a machine-learning classification method combining UML and SML \citep{2022AJ....163...86Z,2023AJ....165...35F} to massive galaxies in the COSMOS-DASH field. Our method gets the sample data completely classified and shows good classification accuracy.

The method includes two steps: (1) UML clustering. In this step, the data is denoised and extracted by CAE. Then the Bagging-based multi-clustering method is used to divide galaxies with similar features into 100 categories at first, and further classified into five categories manually by visual inspection. After discarding sources with inconsistent voting, 47.76\% (8258) of the sources are successfully classified, including 2664 SPHs, 1485 ETDs, 1227 LTDs, 715 IRRs, and 2167 UNCs. (2) SML (i.e., GoogLeNet model) clustering, the 8258 galaxies successfully classified by the UML method are taken as the training set of the GoogLeNet model to train the neural network and successfully classify the remaining 52.24\% of the galaxies. Thus, we achieve the complete morphological classification for our sample.

Our result shows good accuracy in the test set. We also apply the t-SNE graph and $G-M_{20}$ diagram to our classification result, from which we find that the classification results of combining the UML method with the SML method are consistent with the characteristics of the galaxy morphology parameters. 

\begin{acknowledgments}
This paper is based on observations made with the NASA/ESA HST, obtained at the Space Telescope Science Institute, which is operated by the Association of Universities for Research in Astronomy, Inc., under NASA contract NAS 5-26555. These observations are associated with program HSTGO-14114. Support for GO-14114 is gratefully acknowledged. Some of the data presented in this paper were obtained from the Mikulski Archive for Space Telescopes (MAST) at the Space Telescope Science Institute. The specific observations analyzed can be accessed via \dataset[https://doi.org/10.17909/T96Q5M]{https://doi.org/10.17909/T96Q5M}.
This work is supported by the Strategic Priority Research Program of Chinese Academy of Sciences (Grant No. XDB 41000000), the National Natural Science Foundation of China (NSFC, Grant No. 12233008, 11973038, 62106033), the China Manned Space Project (No. CMS-CSST-2021-A07), the Cyrus Chun Ying Tang Foundations and the Frontier Scientific Research Program of Deep Space Exploration Laboratory. C.C.Z. acknowledges the support from Yunnan Youth Basic Research Projects (202001AU070020). Z.S.L. acknowledges the support from the China Postdoctoral Science Foundation (2021M700137). Y.Z.G. acknowledges support from the China Postdoctoral Science Foundation funded project (2020M681281).
\end{acknowledgments}

\bibliographystyle{aasjournal}
\bibliography{ref}
\end{CJK*}
\end{document}